\documentclass[pre, twocolumn]{revtex4-1}
\usepackage{color}
\usepackage{epsfig}
\usepackage{amsmath}
\usepackage{amssymb}

\begin{document}

\title{Phase-field model of vapor-liquid-solid nanowire growth}

\author{Nan Wang$^1$, Moneesh Upmanyu$^2$, and Alain Karma$^1$}
\affiliation{$^1$Physics Department and Center for Interdisciplinary
Research on Complex Systems, Northeastern University, Boston, Massachusetts
02115, USA\\
$^2$Department of Mechanical and Industrial Engineering,
Northeastern University, Boston, Massachusetts 02115, USA}
\email[Corresponding author: Alain Karma, email: ]{a.karma@neu.edu}
\date{\today}

\begin{abstract}
%We present a phase-field model of nanowire growth with vapor-liquid-solid(VLS) growth mechanism based on a multi-phase-field approach where vapor, solid and liquid phase are directly associated with a phase-field parameter. By exploiting the importance of different dynamic time scales at the solid-liquid and the liquid-vapor interfaces, this model recovers the VLS growth theory proposed by Schwarz and Tersoff~\cite{Tersoff} in the sharp-interface limit. Numerical results from this model are carefully benchmarked with the sharp-interface theory. Both size-dependent nanowire growth and size-independent growth are demonstrated within this model. With surface energy anisotropy, profile of the solid-catalyst interface facet matches the sharp-interface prediction and rich growth morphologies are observed.
We present a multi-phase-field model to describe quantitatively
nanowire growth by the vapor-liquid-solid (VLS) process. The
free-energy functional of this model depends on three non-conserved
order parameters that distinguish the vapor, liquid, and solid
phases and describes the energetic properties of various interfaces,
including arbitrary forms of anisotropic $\gamma$-plots for the
solid-vapor and solid-liquid interfaces. The evolution equations for
those order parameters describe basic kinetic processes including
the rapid (quasi-instantaneous) equilibration of the liquid catalyst
to a droplet shape with constant mean curvature, the slow
incorporation of growth atoms at the droplet surface, and
crystallization within the droplet. The standard constraint that the
sum of the phase fields equals unity and the conservation of the
number of catalyst atoms, which relates the catalyst volume to the
concentration of growth atoms inside the droplet, are handled via
separate Lagrange multipliers. An analysis of the model is presented
that rigorously maps the phase-field equations to a desired set of
sharp-interface equations for the evolution of the phase boundaries
under the constraint of force-balance at three-phase junctions
(triple points) given by the Young-Herring relation that includes
torque term related to the anisotropy of the solid-liquid and
solid-vapor interface excess free energies. Numerical examples of
growth in two dimensions are presented for the simplest case of
vanishing crystalline anisotropy and the more realistic case of a
solid-liquid $\gamma$-plot with cusped minima corresponding to two
sets of $(10)$ and $(11)$ facets. The simulations reproduce many of
the salient features of nanowire growth observed experimentally,
including growth normal to the substrate with tapering of the side
walls, transitions between different growth orientations, and
crawling growth along the substrate. They also reproduce different
observed relationships between the nanowire growth velocity and
radius depending on the growth condition. For the basic normal
growth mode, the steady-state solid-liquid interface tip shape
consists of a main facet intersected by two truncated side facets
ending at triple points. The ratio of truncated and main facet
lengths are in quantitative agreement with the prediction of
sharp-interface theory that is developed here for faceted nanowire
growth in two dimensions.
\end{abstract}

\maketitle

\section{Introduction}

Semiconductor nanowires (NWs) have emerged as promising small
building blocks for various nanotechnology applications ranging from
nanoelectronics to sensors to solar energy harvesting. The
functional properties of these NWs can be tuned by controlling their
chemical composition and growth morphologies, and hence a
fundamental understanding of the underlying crystal growth has been
the subject of much recent research
\cite{chemical_review,review_mod_phys,FortunaLi_2010}. A
well-studied NW synthesis route is vapor-liquid-solid (VLS) growth.
In this process, small liquid droplets of a metallic element (e.g.
Au) are deposited on an initially flat solid substrate of the NW
element (e.g. Si). The droplet surfaces act as preferential sites
for the capture of growth atoms by catalytic breakdown of a
molecular gas (e.g. Si$_2$H$_6$). The bulk of the alloy droplet in
turn acts as a conduit of these atoms to the solid, thereby
facilitating the growth of solid regions capped by liquid droplets,
which emerge as NWs from the substrate.

VLS growth has been widely studied experimentally and theoretically
during the past decade. Experiments have revealed a wealth of
interesting growth
behaviors \cite{S.Migration,dia_independent_growth,nucleation_terrace}.
While NWs commonly grow normal to the substrate along some preferred
crystallographic directions (e.g. $\langle 111\rangle$ for Si), they
can also grow at an angle to the
substrate\cite{dia_orientation,orientation_control}, change growth
directions after emerging from the substrate following kinked or
more erratic trajectories, or crawl along the
substrate\cite{drucker_turning}. Furthermore, NWs exhibit intricate
solid-liquid and solid-vapor interface morphologies \cite{Farralis}.
The solid-liquid interface typically consists of a main facet normal
to the NW growth direction, but also small side facets ending at a
triple line where vapor, liquid, and solid phases meet. The
solid-vapor interface that shapes the NW sidewalls consists of
different sets of facets that can be smooth or
sawtooth-like \cite{Sawtooth}. There has been theoretical progress to
address various aspects of NW growth using
analytical  \cite{Givargizov,Schmidt_dia,Lund_diameter_2007,Dubrovskii2008,Roper_voorhees_2007,Roper_voorhees_2010,Asym_unpinning},
numerical continuum
models \cite{Tersoff_PRL2009,Tersoff_NL2011,Tersoff_PRL2011,Tersoff_NL2012,Tersoff_PRL2014,Nan_thesis,PF_NISTa,PF_NISTb,Cai_2014}
and molecular dynamics
simulations \cite{haxhimali_md,moneesh_md,Frolov_md}. Some of those
studies have shed light on the relationship of the NW growth rate
and radius under steady-state growth
conditions \cite{Schmidt_dia,Lund_diameter_2007,Dubrovskii2008}, the
selection of the radius for prescribed liquid catalyst
volume \cite{Roper_voorhees_2007}, the stability and shape changes of
liquid droplets, the oscillatory behavior of side solid-liquid
facets and the growth orientation
selection \cite{Roper_voorhees_2010,PF_NISTb,Asym_unpinning}.

Despite this progress, modeling NW growth quantitatively on
continuum scales remains a major challenge. A main difficulty is the
fact that both solid-liquid and solid-vapor interfaces are faceted
and undergo complex shape changes during NW formation. For example,
as the NW first emerges from the substrate, the equilibrium
solid-liquid interface shape changes from concave to convex and the
solid-vapor interface changes orientation as the NW develops a
tapered shape. Tracking the evolution of faceted solid-liquid and
solid-vapor interfaces under the constraint that different facets
meet with the isotropic liquid-vapor interface at the triple line is
a daunting task. To this end, Schwarz and Tersoff (ST) developed a
continuum model that tracks the evolution of anisotropic non-faceted
interfaces, demonstrating the ability of this model to simulate the
complete evolution of a deposited droplet into a NW
\cite{Tersoff_PRL2009,Tersoff_NL2011,Tersoff_PRL2011,Tersoff_NL2012,Tersoff_PRL2014}.
They further extended this model to faceted interfaces, reproducing
qualitatively non-trivial growth behaviors such as NW kinking and
crawling. Despite its success to reproduce a number of observed
growth modes, this approach relies on phenomenological parameters
and rules to model the energetics and dynamics of facet creation. In
addition, it tracks interfaces explicitly as sharp boundaries, which
makes its extension to 3D difficult.

Phase-field modeling provides an attractive alternative to interface
tracking methods to model VLS growth at the continuum scale. This
method is well-known for its ability to circumvent the difficulty of
interface tracking by making interfaces spatially
diffuse \cite{solidification_review2002,Longqing_chen_microstruture,Steinbach_2009,yuri_nist_pf,yunzhi_wang_defects,Alain_crack,Alain_quantitativePF_PRL,Tomorr_Nature_material}.
Furthermore, by use of more than one scalar order parameters, it can
naturally distinguish several phases, thereby handling complex
geometries and changes of interface topology. However, the
application of the phase-field method to VLS growth remains limited
to date. One set of studies used a phase-field model with viscous
flow to investigate some dynamical aspects of liquid droplet wetting
and shape stability \cite{PF_NISTa,PF_NISTa}. Another study
introduced a multiphase-field model that uses a non-standard hybrid
of Ginzburg-Landau and Cahn-Hilliard dynamics to conserve the volume
of the liquid catalyst \cite{Cai_2014}. This model was used to
produce 3D NW growth morphologies that resemble observed
morphologies. However, the sharp-interface limit of this model was
not analyzed in detail and simulations were carried out for
analytical forms of crystalline anisotropy corresponding to
non-faceted interfaces.

In this paper, we develop a phase-field model to simulate
quantitatively NW growth for realistic forms of crystalline
anisotropy corresponding to faceted interfaces. We carry out a
detailed asymptotic analysis of the model in the limit where the
interface thickness is small compared to the NW radius. This
analysis maps the phase-field dynamical equations onto a
well-defined set of sharp-interface equations, thereby allowing us
to relate phase-field model and materials parameters. Furthermore,
we present 2D simulations that illustrate how the model can
reproduce salient features of NW growth including tapered growth
normal to the substrate, kinking, and crawling. At a more
quantitative level, we validate the model by comparison of 2D
phase-field simulations to predictions of sharp-interface theory.
This comparison is made for the NW growth rate and radius as well as
for the faceted shape of the solid-liquid interface during
steady-state growth at constant velocity.

\subsection{Phase-field formulation}

We develop a multiphase-field formulation with three scalar order
parameters that distinguish the solid, liquid and vapor phases, and
with a free-energy form adapted from a previous model of eutectic
growth \cite{Plapp}. We derive relations that relate this
free-energy functional to energetic properties of various
interfaces, including arbitrary forms of anisotropic $\gamma$-plots
for the solid-vapor and solid-liquid interfaces. The evolution
equations for those order parameters describe basic kinetic
processes including the rapid (quasi-instantaneous) equilibration of
the liquid catalyst to a droplet shape with constant mean curvature,
the catalytic incorporation of growth atoms at the droplet surface,
and crystallization within the droplet. The standard constraint that
the sum of the phase fields equals unity and the conservation of the
number of catalyst atoms, which relates the catalyst volume to the
concentration of growth atoms inside the droplet, are handled via
separate Lagrange multipliers.

Two physically distinct growth situations are modeled. The first is
the one considered by ST where the NW growth rate is limited by the
incorporation rate of growth atoms at the droplet surface. In this
situation, the change of the droplet volume is governed by the
balance of the fluxes of growth atoms into and out of the droplet,
assumed to contain a fixed number of catalyst atoms. During
steady-state growth those two fluxes must balance each other. The
rate of incorporation of growth atoms into the droplet is the
product of the droplet surface area and a constant surface flux $J$,
defined as the number of growth atoms incorporated per unit time per
unit area of droplet surface, while the incorporation rate into the
solid scales as the product of the NW growth rate $V$ and
solid-liquid interface area divided by the atomic volume of solid
$\Omega_s$. Since both the droplet surface area and solid-liquid
interface area scale as $R^{d-1}$ (where $d$ is the spatial
dimension), $V\sim J\Omega_s$ is independent of NW radius in this
limit, as observed in some experiments
\cite{dia_independent_growth}. The second limit we consider is the
one where growth is limited by the solid-liquid interface kinetics
and the chemical potential of growth atoms can be assumed to be
equal in the liquid and vapor and constant in time, i.e. those two
phases equilibrate quickly on the time scale where the solid adds
one additional layer of atoms. In this case, the chemical driving
force for growth (difference of chemical potential between liquid
and solid) is reduced by interface curvature and the growth rate
depends on NW radius as predicted by some sharp-interface theories
and observed in other experiments
\cite{Schmidt_dia,Lund_diameter_2007}.

\subsection{Faceted interfaces}

A general difficulty in modeling faceted interfaces is that the
$\gamma$-plot exhibit cusps at faceted orientations. A cusp is
reflected in the anisotropy of the interface free-energy, which can
be written near a facet in the form
\begin{equation}
\gamma(\theta)=\gamma_f(1+\delta_f |\theta-\theta_f|+\dots) \label{facet}
\end{equation}
where $\theta$ denotes the angle of the normal to the interface with
respect to a fixed reference crystal axis, $\gamma_f$ is the facet
free-energy at angle $\theta=\theta_f$, and the above form is valid
for a vicinal interface where $|\theta-\theta_f|\ll 1$. Such an
interface is generally composed of steps spaced a distance $d\approx
h/|\theta-\theta_f|$ where $h$ is the step height. The excess
interface free-energy associated with step formation is therefore
$\gamma_s/d$, where $\gamma_s$ is the isolated step free-energy
(with unit of energy per unit length of step). It follows that the
total excess free-energy of the vicinal interface is given by Eq.
\ref{facet} with $\delta_f=\gamma_s/(\gamma_fh)$. Cusps make the
function $\gamma(\theta)$ non-differentiable. This poses a
difficulty in phase-field modeling where the evolution equation for
a given phase-field $\phi$ distinguishing two phases is derived from
a first variation of a free-energy functional that contains the
function $\gamma(\hat n)$ with the interface normal expressed as
$\hat n=-\vec\nabla \phi/|\nabla\phi|$. To overcome this difficulty,
we follow the method previously developed for faceted dendritic
solidification that consists of rounding the cusp over a small range
of angle so as to make $\gamma(\theta)$ differentiable
\cite{Jean-Marc}.

In this paper, cusp-rounding is implemented by using the function
$\sqrt{\epsilon^2+x^2}$ that converges to the absolute value
function $|x|$ in the limit $\epsilon\rightarrow 0$. This approach
is conceptually similar to the approach followed in Ref.
\cite{Jean-Marc}, but easier to implement computationally.
Importantly, results do not depend sensitively on cusp-rounding for
small values $\epsilon \sim 10^{-2}$. Furthermore, we consider a
simple form of solid-liquid $\gamma$-plot with two sets of $(10)$
and $(11)$ facets. For a crystal seed surrounded by liquid, this
$\gamma$-plot yields an octagonal faceted equilibrium crystal shape
with four facets of each type. This shape is easily predicted by the
standard Wulff construction. For NW growth from a $(10)$ substrate,
this $\gamma$-plot yields a solid-liquid interface shape consisting
of a main $(10)$ facet intersected by two truncated $(\bar 1 1)$ and
$(11)$ side facets ending at triple points, which is qualitatively
similar to the interface shape observed experimentally and in MD
simulations. In this more complex geometry, the Wulff construction
is not sufficient to predict the interface shape because side facets
end at triple points. To predict this shape, we follow two
equivalent approaches within a sharp-interface picture.

The first approach is to apply the Wulff construction, expressed in
a parametric representation where the cartesian coordinates of the
interface are functions of $\theta$, together with the anisotropic
Young-Herring's condition of thermomechanical equilibrium at triple
points\cite{Herring1951}. This conditions is given by
\begin{equation}
\gamma_{lv} \hat t_{lv} + \gamma_{sl}\hat t_{sl} +\gamma_{sv}\hat t_{sv} + \gamma_{sl}'\hat n_{sl} =0, \label{Young-Herring}
\end{equation}
where $\gamma_{lv}$, $\gamma_{sl}$, and $\gamma_{sv}$ are the
liquid-vapor, solid-liquid, and solid-vapor interfacial energies,
respectively, $\hat t_{\alpha\beta}$ and $\hat n_{\alpha\beta}$ are
the unit vectors tangent and perpendicular to the $\alpha\beta$
interface, respectively, and $\gamma_{sl}'\equiv \partial
\gamma_{sl}/\partial \theta$. This ``torque'' term tends to rotate
the solid-liquid interface towards a low-energy faceted orientation.
A similar term also applies to the anisotropic solid-vapor interface
but is omitted here since it does not influence steady-state NW
growth with vertical side walls. Note that the torque term is
uniquely determined when the cusp is rounded and
$\gamma_{sl}(\theta)$ is differentiable. In this case, the
parametric equation for the interface shape together with Eq.
\ref{Young-Herring} uniquely determines this shape. Importantly,
this shape converges to a unique, physically desired, faceted shape
in the $\epsilon\rightarrow 0$ limit.

The second method to obtain the same shape, which serves as an
independent check, is to treat directly the case $\epsilon=0$
without cusp-rounding where $\gamma_{sl}'$ is not defined at
$\theta=\theta_f$ because of the absolute value in Eq. \ref{facet}.
In this case, the shape is found by considering virtual
displacements of facets and triple points that leave the total
free-energy unchanged. Interestingly, the condition obtained by
considering the virtual displacement of a triple point can also be
obtained by projecting the Young-Herring condition
(\ref{Young-Herring}) onto two cartesian axes parallel and
perpendicular to the NW growth direction, which yields two
equations. The solutions of those two equations for fixed faceted
orientations in turn yield the value undetermined of $\gamma_{sl}'$
at $\theta=\theta_f$, which must physically be comprised in the
interval $-\gamma_f\delta_f<\gamma_{sl}'<\gamma_f\delta_f$ following
Eq. \ref{facet}, and a second equation identical to the one obtained
by considering a virtual displacement of the triple point that
extends the truncated side facet of the solid-liquid interface. This
latter condition determines the dihedral angle between the
liquid-vapor and solid-liquid interfaces at the triple point. For
this reason, in the $\epsilon\rightarrow 0$ limit, the rounded cusp
treatment yields the same shape as the one obtained by considering
virtual displacements of facets and triple points. We find that this
faceted shape predicted by sharp-interface theory is in good
quantitative agreement with the one obtained by phase-field
simulations. Therefore, in addition to validating our phase-field
approach, our results also clarify how sharp-interface theory should
be formulated to predict faceted interface shapes during NW growth.

\subsection{Outline}

This paper is organized as follows. In the next section, we write
down the set of sharp-interface equations used to describe NW
growth, which follow closely the model introduced by Schwarz and
Tersoff \cite{Tersoff_PRL2009}. In section III, we present our multiphase-field model,
which is formulated to reduce to the sharp-interface equations of
section II. Various ingredients of the model including the
free-energy landscape, the description of the driving force for NW
growth, the treatment of Lagrange multipliers to satisfy constraints
imposed on the droplet volume and the sum of phase fields, the
evolution of the concentration within the droplet, interface
mobility, and interface free-energy, are summarized in separate
subsections. The equations of the model are then summarized followed
by a description of the treatment of anisotropic interfaces in the
last subsection. Next, in section IV, we analyze the sharp-interface
limit of the model. This analysis is used to pinpoint the conditions
under which this limit reduces to the desired set of sharp-interface
equations and to relate phase-field parameters to materials
parameters. Various numerical examples are then presented in section
V. The dependence of the NW growth velocity on radius is
characterized in different limits for isotropic interface energies
and simulations of faceted growth are compared to the predictions of
sharp-interface theory. Conclusions  are presented in the last
section.

\section{Sharp-interface model}
VLS NW growth involves 3 steps essentially~\cite{Schmidt_dia}: (1)
incorporation of Si precursors from the vapor at the vapor-catalyst
interface, (2) diffusion of Si atoms through Au catalyst droplet and
(3) crystallization at the solid-liquid interface. The ST
sharp-interface (SI) model~\cite{Tersoff_PRL2009} ignores the
diffusion process in step (2) and uses a uniform chemical potential
within the catalyst droplet since diffusion through the nanoscale
liquid droplet is fast compared to the NW growth rate.

The crystallization in step (3) is driven by over-saturation of Si
atoms in the catalyst. The NW growth rate $v_n$ is related to the
difference of chemical potential between the solid Si and liquid
catalyst as
\begin{equation}
v_n=\frac{M_{sl}}{\Omega_s}(\mu_l-\mu_s) \label{sl_growth_law},
\end{equation}
assuming a linear dependence. Here, $M_{sl}$ is the solid-liquid
interface mobility, $\Omega_s$ is the atomic volume of Si atoms in
the solid. The liquid concentration is given by $c_l=N_g/(N_c+N_g)$
where $N_g$ is the number of Si growth atoms and $N_c$ the number of
catalyst atoms. The chemical potential in the liquid is assumed to
be directly related to the over-saturation $c_l-c_0$ as
\begin{equation}
\mu_l=\beta(c_l-c_0)+\Omega_l\gamma_{lv}\kappa_{lv},\label{sl_mul}
\end{equation}
where $c_0$ is the equilibrium concentration and
the second term on the right-hand-side is the Gibbs-Thomson correction related to the curvature $\kappa_{lv}$ of the liquid-vapor surface with energy $\gamma_{lv}$.
$\Omega_l$ is the atomic volume of Si atoms in the liquid and
$\beta\equiv \partial \mu_l/\partial c$. The chemical potential in the solid can be written as
\begin{equation}
\mu_s=\Omega_s\left[\left(\gamma_{sl}+\frac{d^2\gamma_{sl}}{d\theta^2
}\right)\kappa_{sl}+p\right].\label{sl_mus}
\end{equation}
The first term is the generalized Gibbs-Thomson effect for an anisotropic solid-liquid interfacial free-energy $\gamma_{sl}(\theta)$
where $\theta$ is the local surface orientation angle and $\kappa_{sl}$ is
the solid-liquid interface curvature. The second term is a normal
force on the solid coming from the liquid internal pressure
$p=\gamma_{lv}\kappa_{lv}$.  In addition, thermo-mechanical equilibrium at the triple junction of the three (solid, liquid, and vapor)
phases imposes a geometrical constraint on the dihedral angles between the phase boundaries at this junction, which is given by the anisotropic Young-Herring's condition (Eq. \ref{Young-Herring}).

To calculate the liquid concentration, the number of Si atoms $N_g$
in the catalyst is tracked during the growth. Assuming a constant
flux $J$ on the catalyst surface (the number of Si atoms
incorporated at the liquid-vapor surface per unit area per unit
time), we have
\begin{equation}
dN_g/dt=\int_{lv}J ds-\Omega_s^{-1}\int_{sl} v_n
ds,\label{growth_si_dndt}
\end{equation}
where $ds$ is the surface element, and $v_n$ is the NW growth
velocity. The first integral covers the liquid-vapor surface and
accounts for the Si source flux in step (1), the second integral
covers the solid-liquid surface and serves as a Si atoms sink due to
the crystallization in step (3). During steady-state growth,
$dN_g/dt$ in Eq. \ref{growth_si_dndt} vanishes and the two
fluxes balance each other.

%If the catalyst size can change through surface migration or other
%mechanisms as seen in some experiments, radius of the growing wire
%may change with the catalyst size as a result of Young's
%condition\cite{S.Migration}. For some combinations of the surface
%energies, the Young-Herring's condition simply can not be satisfied
%if the solid-vapor side wall is perpendicular to the substrate.
%Re-positioning of the catalyst to a stable site may happen in that
%case and induces NW growth in other directions.

\section{Phase-field model}

\subsection{Multiphase-field formulation}
\begin{figure}[!h]
\includegraphics[width=85mm]{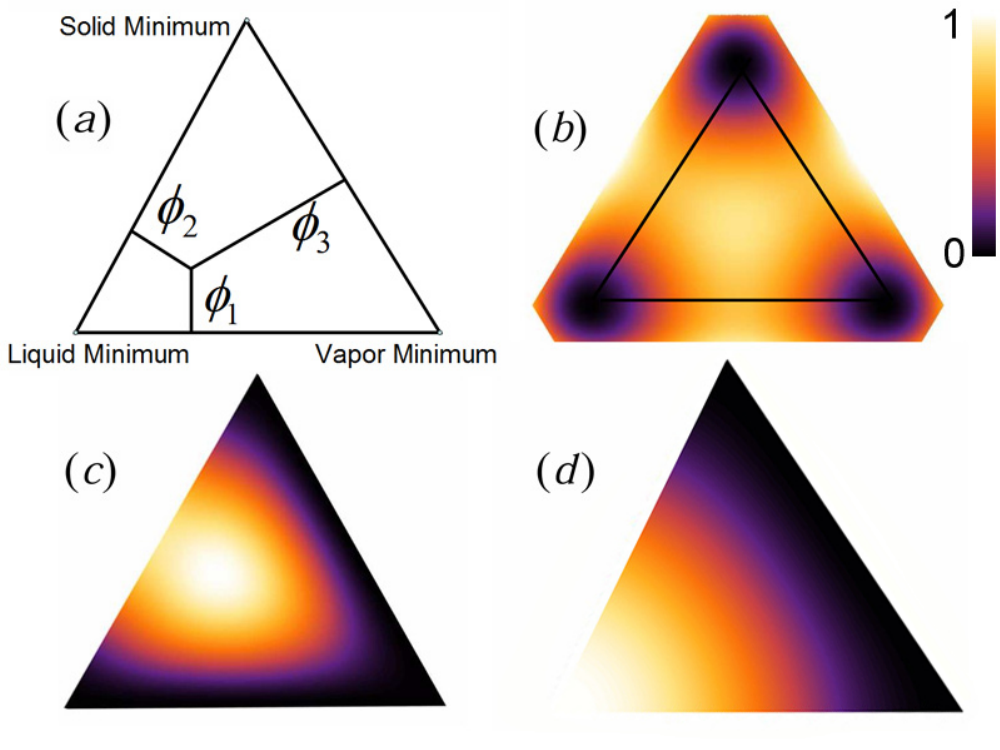}
\caption{Basic features of the multiphase-field model used in our
formulation. (a) Multiphase-field parameters in a Gibbs triangle.
Three bulk phases occupy the 3 vertices of the triangle. The 3 edges
correspond to the 3 binary interfaces in the model. A point in the
triangle is given by $(\phi_1,\phi_2,\phi_3)$ where $\phi_i$ is the
distance from the given point to the $j-k$ binary interface. (b)
Free-energy landscape of the potential term $f_p$ in Eq.~\ref{pf_f_p}
with $a_i=b_i=0$. To highlight the three free-energy minima, the plot
range of  $(\phi_1,\phi_2,\phi_3)$ is extended beyond [0,1] and thus outside the Gibbs triangle in (a) reproduced as a thick black line.
(c) Free-energy
landscape of the $a_if_a^i+b_if_b$ part in Eq.~\ref{pf_f_p}. By
setting a non-zero $a_2$, it forms an additional barrier along the
solid-liquid binary interface and leaves the other two binary
interfaces unchanged. $b_i$ controls the height of the triple point
$(\phi_1=\phi_2=\phi_3=1/3)$. (d) The function $g_l$ given in
Eq.~\ref{function_g_l}. It equals to $1$ in the liquid and smoothly
decreases to $0$ at the solid-vapor binary interface.}
\label{mpf_basic}
\end{figure}

To model the VLS NW growth within the PF framework, we build our
model on the well established multiphase-field approach developed in
the context of multiphase solidification~\cite{Plapp}. We use three
order parameters $\phi_1$, $\phi_2$ and $\phi_3$ to distinguish the solid, vapor, and liquid phases, respectively.
$\phi_i$
is the fraction of the $i^{th}$ phase at a given point in space (as
depicted in Fig.~\ref{mpf_basic}(a)) with constraints
$\phi_i\in[0,1]\vee i$ and
\begin{equation}
\sum_{i=1}^3 \phi_i=1 \label{constraint_1}.
\end{equation}

%The NW growth front i.e. the solid-liquid interface is then
%characterized by a steep but smooth variation of the solid and
%liquid parameter ($\phi_1$ and $\phi_3$) as shown in
%Fig. \ref{pf_demo}. The other two-phase interfaces are also captured
%in similar way (Fig.\ref{pf_demo}).
\begin{figure}[!h]
\includegraphics[width=85mm]{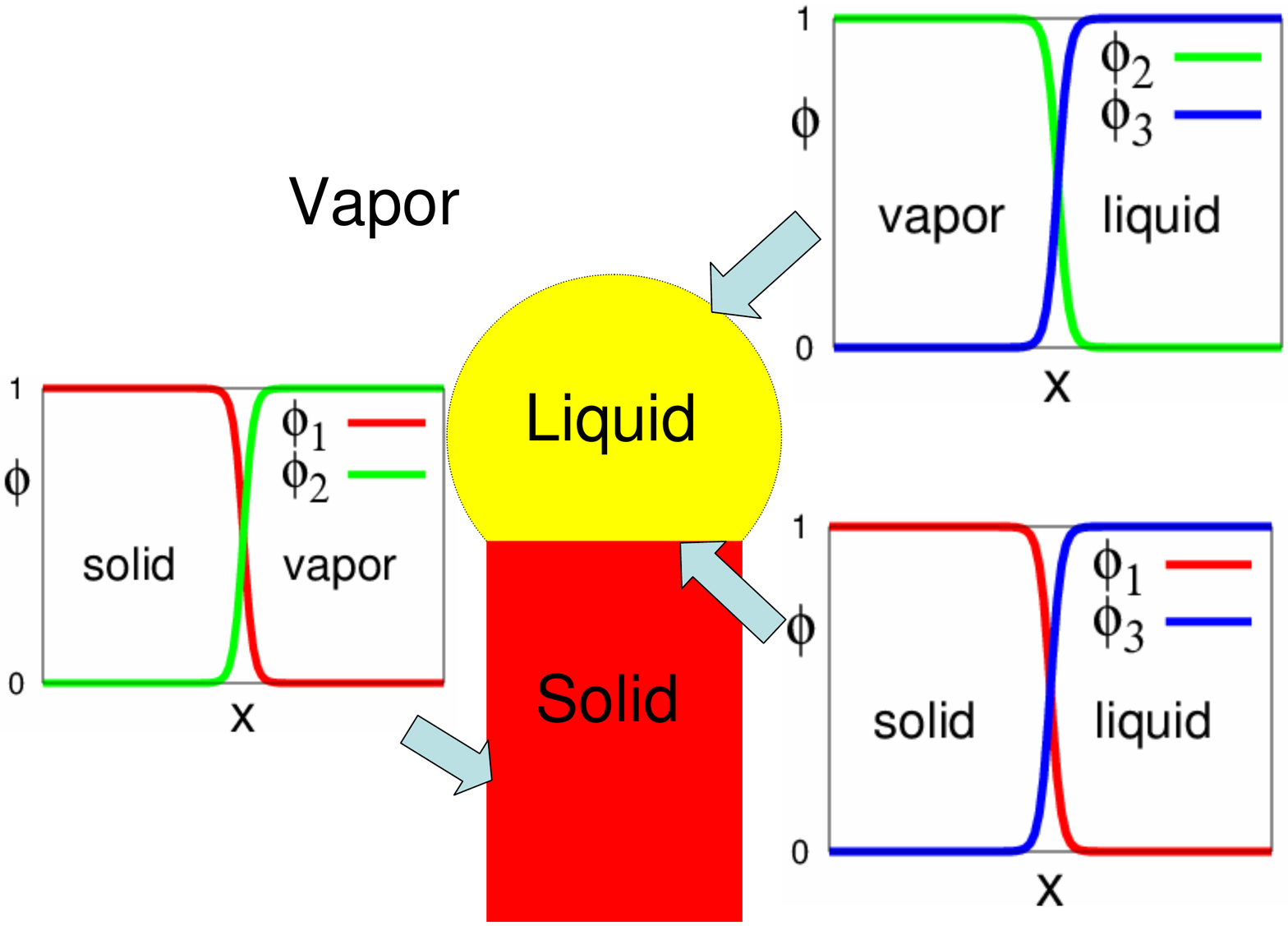}
\caption{Outline of the 3 binary interfaces involved in VLS NW
growth. The phase fields vary
smoothly from 0 to 1 across a spatially diffused interface region with
$\phi_1=1$, $\phi_2=1$, and $\phi_3=1$ in solid, vapor, and liquid, respectively.}
\label{pf_demo}
\end{figure}

The Lyapounov functional representing the total free-energy of this multiphase system is chosen to have a similar form as the Folch-Plapp model of eutectic solidification \cite{Plapp}
\begin{equation}
F=\int \left(\frac{\sigma}{2}f_k+hf_p\right)dv,\label{pf_energy_iso}
\end{equation}
where $\sigma$ and $h$ are dimensional constants and $dv$ is the volume
element. The first term
\begin{equation}
f_k=\sum_{i=1}^3|\nabla\phi_i|^2
\end{equation}
describes the free-energy cost associated with the spatial variation of the
phase-field within the interface regions. The second term
\begin{equation}
f_p=\sum_{i=1}^3[f_d^i+a_if_a^i+b_if_b],\label{pf_f_p}
\end{equation}
corresponds to the bulk free-energy density with
$f_d^i=\phi_i^2(1-\phi_i)^2$,
$f_a^i=\phi_j^2\phi_k^2(2\phi_j\phi_k+3\phi_i)$ and
$f_b=\phi_i^2\phi_j^2\phi_k^2$. This form yields 3 free-energy
minima for the bulk phases with tunable inter-phase barrier heights
(Fig.~\ref{mpf_basic}(b)). By choosing different $a_i$, the height
of the free-energy barrier between phase j and k can be modified,
while leaving the i-j and i-k barriers unaffected. Since this
free-energy barrier is linked to the interface energy in the PF
model, one can tune the free-energy of any binary interface using
this function as shown in later sections. $b_i$ in the last term is
multiplied by the square of all three phases. It can be used to vary
the height of the three phase junction (triple-point) region in the
free-energy landscape (Fig.~\ref{mpf_basic}(c)), with the constraint
$b_i>9a_i/2$ ensuring that the triple point is not energetically
favored over a binary interface \cite{Plapp}. Increasing $b_i$
decreases the size of the triple-point region. Different choices of
this parameter only affect the model behavior within this region but
does not change the sharp-interface limit of the model discussed in
section~\ref{growth_si}.

The advantage of the phase-field model is that it satisfies automatically the anisotropic
Young-Herring condition that does not need to be imposed as an additional constraint.
However, to track the NW growth, a few other conditions also need to be taken
into account. In addition to the constraint on the phase fields $\sum_{i=1}^3
\phi_i=1$, the catalyst
volume change needs to be considered during the NW growth. To
incorporate that, a modified functional with a Lagrange multiplier
$\lambda_A$ is introduced
\begin{equation}
\tilde{F}=F-\lambda_Ah\left[\int
g_l(\vec{\phi})dv-A(t)\right].\label{func_vconstraint}
\end{equation}
$g_l(\vec{\phi})$ is a function that varies smoothly from 1 in the
liquid to 0 in the other two phases (Fig.~\ref{mpf_basic}(d)) such
that its integral over space can be used as a measure of the
catalyst size. This additional term ensures that, at given time $t$,
the catalyst area in 2D  (volume in 3D) is given by $A(t)$. Specifically, the liquid tilt
function $g_l$ is chosen to be
\begin{eqnarray}
g_l&=&\frac{\phi_3^2}{4}\left\{ 15(1-\phi_3)[1+\phi_3-(\phi_2-\phi_1)^2]\right. \label{function_g_l}\\
\nonumber && \left.+ \phi_3(9\phi_3^2-5) \right\}.
\end{eqnarray}
%The form of $A(t)$ can either be controlled by a different physics
%or be set to a constant in case of insignificant volume change.

From the modified energy functional $\tilde{F}$, evolution equations
for $\phi_i$ are derived in a variational form
\begin{equation}
\tau\frac{\partial \phi_{i}}{\partial
t}=-\frac{K(\vec{\phi})}{h}\frac{\delta \tilde{F}}{\delta \phi_{i}},
\label{growth_vonly_eqn}
\end{equation}
where $\tau$ is a relaxation time constant  and $K(\vec{\phi})$ is a function that can be directly related to
the sharp-interface mobility (section \ref{growth_si}).

\subsection{Driving force for crystallization}
The variational formulation above is just a commonly used scheme to
derive PF evolution equations that drive a multiphase system towards
a global free-energy minimum. For VLS growth, however, the growth is
externally driven by a flux of precursor atoms incorporated at the
liquid-vapor interface. This flux maintains a finite supersaturation
in the droplet. This supersaturation in turn drives the
crystallization of growth atoms at the solid-liquid interface. To
account for the fact that this driving force is localized at the
solid-liquid interface, we add to the right-hand-side of Eq.
\ref{growth_vonly_eqn} a term $\Delta\mu \Omega_s^{-1}
u(\phi_1,\phi_3)$ for the evolution of $\phi_1$ and $\phi_3$ only.
Physically, this additional term can be interpreted as the
difference of chemical potential of Si atoms between the NW solid
and liquid catalyst denoted here by $\Delta\mu$ where the additional
factor of $\Omega_s^{-1}$ converts the unit of energy per atom of
the chemical potential to energy per volume in the PF model. We will
derive later an expression for $\Delta\mu$ from the requirement that
the evolution of the growth atom concentration in the droplet is the
same in the phase-field and sharp-interface models. To localize the
driving force for NW growth at the solid-liquid interface,
$\Delta\mu$ is multiplied by the function
$u(\phi_1,\phi_3)=15\phi_1^2\phi_3^2$. On the binary solid-liquid
interface, where $\phi_3=1-\phi_1$ and $\phi_2=0$, this term can be
derived variationally from the bulk free-energy density term $-\int
d\phi_1u(\phi_1,1-\phi_1)$, which is the standard quintic polynomial
used in phase-field models of monophase solidification that lowers
the free-energy of the solid $(\phi_1=1)$ with respect to the liquid
($\phi_1=0$) \cite{Alain_quantitativePF_PRE}. The same quintic
polynomial is used here to provide the driving force for NW growth.
However, because the driving force must be physically localized at
the solid-liquid interface, i.e. there is no driving force for the
solid to grow into the vapor phase, the $\Delta\mu \Omega_s^{-1}
u(\phi_1,\phi_3)$ term is only added to the evolution equation of
$\phi_1$ and $\phi_3$ but not $\phi_2$. This procedure makes the
phase-field model globally non-variational since multiphase-field
equations with a driving force localized on a specific binary interface
cannot be derived in the standard variational form of Eq.
\ref{growth_vonly_eqn} by simply adding a free-energy contribution
to $F$ that depends on $\phi_1$, $\phi_2$, and $\phi_3$.  However, previous
studies have shown that non-variational and variational phase-field
formulations are equally well-suited to model dendritic
solidification as long as they reduce to the desired sharp-interface
limit \cite{Dantzig1999}. Numerical examples presented in the results section confirm
that the non-variational multiphase-field formulation developed here
yields physically realistic NW growth behaviors that can be
quantitatively related to sharp-interface theory.

The evolution equations that incorporate the catalyst volume change
and this localized driving force for crystallization can be written
in the compact form
\begin{equation}
\tau\frac{\partial \phi_{i}}{\partial
t}=-\frac{K(\vec{\phi})}{h}\frac{\hat{\delta}
\tilde{F}}{\hat{\delta} \phi_{i}},
\end{equation}
with
\begin{equation}
\frac{\hat{\delta} \tilde{F}}{\hat{\delta}
\phi_{i}}= \frac{\hat{\delta} F}{\hat{\delta}
\phi_{i}}-\lambda_Ah\frac{\partial g_l}{\partial \phi_i},
\end{equation}
which incorporates the change of catalyst volume, and
\begin{equation}
\frac{\hat{\delta} F}{\hat{\delta} \phi_{1}}=\frac{\delta
F}{\delta \phi_{1}}-\Delta\mu\Omega_s^{-1}
u(\phi_1,\phi_3),\label{del_F_hat1}
\end{equation}
\begin{equation}
\frac{\hat{\delta} F}{\hat{\delta} \phi_{2}}=\frac{\delta F}{\delta
\phi_{2}},\label{del_F_hat2}
\end{equation}
\begin{equation}
\frac{\hat{\delta} F}{\hat{\delta} \phi_{3}}=\frac{\delta
F}{\delta \phi_{3}}+\Delta\mu\Omega_s^{-1}
u(\phi_1,\phi_3),\label{del_F_hat3}
\end{equation}
which incorporate the driving force for crystallization only at the solid-liquid interface.

\subsection{Lagrange multipliers and constraints}
The equations above control the catalyst volume through the Lagrange
multiplier $\lambda_A$ but have not yet included the phase fraction
condition $\sum_{i=1}^3 \phi_i=1$. It has been shown previously
\cite{Plapp} that such a condition is satisfied by writing the
equations of motion as
\begin{equation}
\tau\frac{\partial \phi_{i}}{\partial
t}=-\frac{K(\vec{\phi})}{h}\left(\frac{\hat{\delta}
\tilde{F}}{\hat{\delta}
\phi_{i}}-\frac{1}{3}\sum_{i=1}^3\frac{\hat{\delta}
\tilde{F}}{\hat{\delta} \phi_{i}}\right) {\rm for} \;
i=1,2,\label{growth_final_eqn}
\end{equation}
and
\begin{equation}
\phi_3=1-\phi_1-\phi_2.\label{growth_final_eqn_fraction}
\end{equation}
The expression of the Lagrange multiplier $\lambda_A$ derived in appendix \ref{growth_appendix_lag} is given by
\begin{widetext}
\begin{equation}
\lambda_A=\frac{\int Kh^{-1}\sum_{i=1}^2 \frac{\hat{\delta}
F}{\hat{\delta} \phi_i}\frac{\partial \tilde{g_l}}{\partial
\phi_i}dv- \frac{1}{3}\int Kh^{-1}\sum_{i=1}^2 \frac{\partial
\tilde{g_l}}{\partial \phi_i}\sum_{j=1}^3\frac{\hat{\delta}
F}{\hat{\delta} \phi_j}dv+\dot{A}\tau}{\int
K\sum_{i=1}^2\left(\frac{\partial
\tilde{g_l}}{\partial\phi_i}\right)^2 dv-\frac{1}{3}\int
K\left(\sum_{i=1}^2 \frac{\partial \tilde{g_l}}{\partial
\phi_i}\right)^2dv},\label{lambda_v_main}
\end{equation}
\end{widetext}
where $\tilde{g_l}$ is a modified $g_l$ function with $\phi_3$
replaced by $1-\phi_1-\phi_2$, and $\dot{A}\equiv dA/dt$
is related to the evolution of the droplet concentration in the next subsection.

\subsection{Droplet concentration evolution}
The driving force $\Delta\mu$ at the solid-liquid interface
generally depends on the over-saturation in the catalyst and the
liquid-vapor interface curvature. To calculate the over-saturation
part, a PF analog of Eq. \ref{growth_si_dndt} for the evolution of the number of growth
atoms $N_g$ in the catalyst needs to be included in the model. In
terms of the PF variables, this evolution equation can be expressed as
\begin{equation}
\frac{dN_g}{dt}=\frac{J}{\eta}\int \phi_2\phi_3
dv-\frac{1}{\Omega_s}\int
\partial_tg_s(\vec{\phi})dv,\label{pf_dndt}
\end{equation}
where $\eta$ in the first flux is a normalization constant with unit
of length chosen such that $\eta^{-1}\int dv\phi_2\phi_3=\int_{lv}
ds$ in the sharp-interface limit. Using a solid measuring function
(similar to $g_l$ in measuring liquid size)
\begin{eqnarray}
g_s&=&\frac{\phi_1^2}{4}\left\{ 15(1-\phi_1)[1+\phi_1-(\phi_3-\phi_2)^2]\right. \\
\nonumber && \left.+ \phi_1(9\phi_1^2-5)
\right\},\label{coupling_gs}
\end{eqnarray}
time derivative of the $g_s$ integral (the second term on the
right-hand-side of Eq. \ref{pf_dndt}) corresponds to the increasing
rate of solid (i.e. the crystallization of Si atoms). During
steady-state growth those two fluxes must balance each other. With
$c_l=N_g/(N_c+N_g)$, the contribution of the over-saturation to the
driving force is then given by $\beta(c_l-c_0)$, and the total
driving force can be written as
\begin{equation}
\Delta\mu=-[\beta(c_l-c_0)-\lambda_Ah\Omega_l],\label{pf_mu}
\end{equation}
where the first term is the over-saturation contribution and the
second term is the Lagrange multiplier which can be reduced to the
liquid-vapor curvature effect as shown later in the sharp-interface
asymptotics. The droplet volume $A=\Omega_lN_g+A_c$ where $A_c$ is
the volume contribution of catalyst atoms assumed to remain
constant. The rate of change of the volume in Eq.
\ref{lambda_v_main} is then given by
\begin{equation}
\dot{A}=\Omega_l\frac{dN_g}{dt},\label{pf_dvdt}
\end{equation}
where $dN_g/dt$ is given by Eq. \ref{pf_dndt}.
Also, at any time, the catalyst volume can be expressed as
\begin{equation}
A=A_0\frac{1-c_0}{1-c_l},\label{v_change_analytic}
\end{equation}
where $A_0$ is the catalyst size at $c_l=c_0$. In the limit of small
supersaturation ($c_l$ close to $c_0$), $A\approx A_0$.

\subsection{Interface mobility}

While the liquid-vapor interface relaxes rapidly to a shape with
constant mean curvature, NW growth is a comparatively much slower
process controlled in different limits by the crystallization
kinetics or the incorporation of Si atoms from the vapor phase. In
addition, the evolution of the solid-vapor interface by surface
diffusion is essentially frozen on the time scale of NW growth.
Accordingly, the mobility function $K(\vec{\phi})$ should be chosen
such that the solid-vapor interface mobility vanishes far from the
triple junctions while the liquid-vapor interface mobility $M_{lv}$
is much larger than the solid-liquid interface mobility $M_{sl}$. We
use the form
\begin{equation}
K(\vec{\phi})=(1-4\phi_1\phi_2)(1+\alpha\phi_2\phi_3),\label{growth_mob_K}
\end{equation}
where $\alpha$ is chosen such that
\begin{equation}
\frac{\int_0^1
\phi_2(1-\phi_2)\sqrt{1+a_2\phi_2(1-\phi_2)}d\phi_2}{\int_0^1
\phi_1(1-\phi_1)\sqrt{1+a_1\phi_1(1-\phi_1)}K^{-1}
d\phi_1}=M_{lv}/M_{sl}.\label{growth_mob_ratio}
\end{equation}
This integral condition will become clearer after we derive the
sharp-interface limit of the PF equations.

\subsection{Interface free-energy}
The three binary interfaces involved in NW growth generally have
different excess free-energy.
The excess free-energy of the interface between phases where
$\phi_j=1$ and $\phi_k=1$ is given by an integral of the free-energy density across the
interface~\cite{Plapp}
\begin{equation}
\gamma_{jk}=2\sqrt{2}Wh\int_0^1
p(1-p)\sqrt{1+a_ip(1-p)}dp,\label{gamma_general}
\end{equation}
where $W=\sqrt{\sigma/h}$, $p$ is either $\phi_j$ or $\phi_k$ (since
$\phi_j=1-\phi_k$ along the j-k interface) and $a_i$ is the
coefficient appearing in Eq. \ref{pf_f_p}. This expression can be used to
incorporate experimentally relevant surface energies for the
solid-liquid, solid-vapor, and liquid-vapor interfaces into this model
by using different coeffficients $a_i$, $a_j$ and $a_k$ to vary the
free-energy barriers between the j-k, i-k, and i-j phases, respectively.
The expression above is limited to isotropic interfaces and
the incorporation of crystalline anisotropy will be discussed later on.

\subsection{Summary of phase-field model equations}

We summarize here the PF model equations presented in the previous
subsections. Even though some formulae are derived in the subsequent
sharp-interface analysis, this self-contained summary is intended to
facilitate the numerical implementation of the model. In the ST
sharp-interface model, NW growth is governed by the growth law of
Eq. \ref{sl_growth_law} and the droplet concentration evolution of
Eq. \ref{growth_si_dndt}. In the PF model, the growth law is
embodied in Eqs. \ref{growth_final_eqn} and
\ref{growth_final_eqn_fraction}, which can be explicitly written as
\begin{eqnarray}
\nonumber\tau\frac{\partial \phi_1}{\partial
t}&=&K(\vec{\phi})\Bigg[W^2\nabla^2\phi_1-2\phi_1(1-\phi_1)(1-2\phi_1)\\
\nonumber&&-\sum_{i=1}^3\frac{\partial (a_if_a^i+b_if_b)}{\partial
\phi_1}+\Delta\tilde{\mu}u(\phi_1,\phi_3)+\lambda_A\frac{\partial
g_l}{\partial
\phi_1}-\frac{1}{3}\lambda_{\phi}\Bigg],\\
\label{iso_eqn1}
\end{eqnarray}
\begin{eqnarray}
\nonumber\tau\frac{\partial \phi_2}{\partial
t}&=&K(\vec{\phi})\Bigg[W^2\nabla^2\phi_2-2\phi_2(1-\phi_2)(1-2\phi_2)\\
\nonumber&&-\sum_{i=1}^3\frac{\partial (a_if_a^i+b_if_b)}{\partial
\phi_2}+\lambda_A\frac{\partial g_l}{\partial
\phi_2}-\frac{1}{3}\lambda_{\phi}\Bigg],\\
\label{iso_eqn2}
\end{eqnarray}
\begin{equation}
\phi_3=1-\phi_1-\phi_2,\label{iso_eqn3}
\end{equation}
with $u(\phi_1,\phi_3)=15\phi_1^2\phi_3^2$,
\begin{equation}
\Delta\tilde{\mu}=\Delta\mu/(h\Omega_s),
\end{equation}
\begin{eqnarray}
\nonumber\lambda_{\phi}&=&\sum_{j=1}^3\bigg[W^2\nabla^2\phi_j-2\phi_j(1-\phi_j)(1-2\phi_j)\\
&&-\sum_{i=1}^3\frac{\partial (a_if_a^i+b_if_b)}{\partial
\phi_j}+\lambda_A\frac{\partial g_l}{\partial \phi_j}\bigg],
\end{eqnarray}
and
\begin{equation}
\lambda_A=\frac{I_1- I_2+\dot{A}\tau}{I_3-I_4},\label{lag_sum_I}
\end{equation}
where
\begin{equation}
I_1=\int Kh^{-1}\sum_{i=1}^2 \frac{\hat{\delta} F}{\hat{\delta}
\phi_i}\frac{\partial \tilde{g_l}}{\partial \phi_i}dv,\label{int_I1}
\end{equation}
\begin{equation}
I_2=\frac{1}{3}\int Kh^{-1}\sum_{i=1}^2 \frac{\partial
\tilde{g_l}}{\partial \phi_i}\sum_{j=1}^3\frac{\hat{\delta}
F}{\hat{\delta} \phi_j}dv,\label{int_I2}
\end{equation}
\begin{equation}
I_3=\int K\sum_{i=1}^2\left(\frac{\partial
\tilde{g_l}}{\partial\phi_i}\right)^2 dv,\label{int_I3}
\end{equation}
\begin{equation}
I_4=\frac{1}{3}\int K\left(\sum_{i=1}^2 \frac{\partial
\tilde{g_l}}{\partial \phi_i}\right)^2dv,\label{int_I4}
\end{equation}
with $(\hat{\delta} F)/(\hat{\delta} \phi_i)$ defined by
Eqs. \ref{del_F_hat1}, \ref{del_F_hat2}, and \ref{del_F_hat3}, and
$\tilde{g_l}$ obtained by replacing $\phi_3$ with $1-\phi_1-\phi_2$ in
Eq. \ref{function_g_l}. As we will see later in the sharp-interface
analysis, the equations above can recover exactly the growth
law in the ST model (Eq.\ref{sl_growth_law}). The droplet concentration
evolution in the PF model is given by Eq.\ref{pf_dndt}
\begin{equation}
\frac{dN_g}{dt}=\frac{J}{\eta}\int \phi_2\phi_3
dv-\frac{1}{\Omega_s}\int
\partial_tg_s(\vec{\phi})dv,
\end{equation}
where $J$ is the incorporation flux at the liquid-vapor interface,
$\eta$ is a constant with unit of length chosen such that
$\eta^{-1}\int dv\phi_2\phi_3$ gives the liquid-vapor surface area,
$g_s$ is the function given in Eq.~\ref{coupling_gs}. With this
definition $\eta=\int dn \phi_2(n)\phi_3(n)$ where $n$ is the
coordinate normal to the liquid-vapor interface and $\phi_2$ and
$\phi_3=1-\phi_2$ are the stationary one-dimensional phase-field
profiles corresponding to an equilibrium interface. The value of
$\eta$ is given in the numerics section. In addition,
$\Delta\tilde{\mu}$ is related to the Si atom concentration in the
droplet $c_l=N_g/(N_c+N_g)$ and $\lambda_A$ by
\begin{equation}
\Delta\tilde
\mu=-\left[\frac{\beta(c_l-c_0)}{h\Omega_s}-\lambda_A\frac{\Omega_l}{\Omega_s}\right],
\end{equation}
and the catalyst volume evolution is
included by replacing $\dot{A}$ in
Eq. \ref{lag_sum_I} with Eq. \ref{pf_dvdt}
\begin{equation}
\frac{dA}{dt}=\Omega_l\frac{dN_g}{dt}.
\end{equation}
The interfacial free-energies are
given by
\begin{equation}
\gamma_{sl}=2\sqrt{2}Wh\int_0^1
p(1-p)\sqrt{1+a_2p(1-p)}dp,\label{G_sl_eqn_sum}
\end{equation}
\begin{equation}
\gamma_{sv}=2\sqrt{2}Wh\int_0^1 p(1-p)\sqrt{1+a_3p(1-p)}dp,
\end{equation}
\begin{equation}
\gamma_{lv}=2\sqrt{2}Wh\int_0^1
p(1-p)\sqrt{1+a_1p(1-p)}dp.\label{G_lv_eqn_sum}
\end{equation}
The solid-liquid interface mobility is
\begin{equation}
M_{sl}=\frac{W^2}{\tau \gamma_{sl}}.\label{M_sl_eqn_sum}
\end{equation}
The last parameter we need to specify is related to the implicit
assumption in the sharp-interface model that the liquid-vapor interface relaxes quasi-instantaneously to an
equilibrium shape on the time scale of NW growth. This limit can be modeled by choosing
$\alpha$ in the expression for the mobility
\begin{equation}
K(\vec{\phi})=(1-4\phi_1\phi_2)(1+\alpha\phi_2\phi_3),\label{K_mob_eqn_sum}
\end{equation}
such that the condition
\begin{equation}
\frac{\int_0^1 p(1-p)\sqrt{1+a_2p(1-p)}dp}{\int_0^1
p(1-p)\sqrt{1+a_1p(1-p)}K^{-1} dp}=\frac{M_{lv}}{M_{sl}}\gg1
\end{equation}
is satisfied where $K=1+\alpha p(1-p)$ in the above expression where
$K(\vec{\phi})$ is evaluated at the liquid-vapor interface with
$\phi_1=0$, $\phi_2=p$, and $\phi_3=1-p$, and the value of $\alpha$
is given in the numerical implementation part in
section~\ref{num_results}.

\subsection{Incorporation of crystalline anisotropy and facets}
\label{secanis}

The incorporation of crystalline anisotropy in monophase
solidification models has been treated in various studies for
atomically rough interfaces without
\cite{kobayashi_aniso,mcfadden_aniso} and with \cite{sharp_corner1}
missing orientations as well as for faceted interfaces
\cite{Jean-Marc}. To incorporate the anisotropy of the solid-liquid
interface excess free-energy in a multiphase system, an extension to
the original Folch-Plapp model \cite{Plapp} needs to be developed.
Because of the phase fraction condition $\sum_i\phi_i=1$,
incorporation of anisotropy through orientation-dependent gradient
terms, as in monophase solidification, is problematic. Therefore, we
follow an alternate approach that consists of making the free-energy
barrier height between the solid and liquid phases
orientation-dependent. Details of this approach are given in
appendix \ref{growth_appendix_facet} and only the main results are
summarized here. The functional derivatives are modified as
\begin{eqnarray}
\frac{1}{h}\frac{\delta F}{\delta \phi_i}&=&\frac{\partial
f_d^i}{\partial
\phi_i}-W^2\nabla^2\phi_i+\sum_{l=1}^3\Bigg[a_l\frac{\partial
f_a^l}{\partial \phi_i}+b_l\frac{\partial f_b}{\partial
\phi_i}\label{facet_fd1}\\
\nonumber&&+\frac{\partial}{\partial
x}\left(\frac{\phi_{i,y}}{|\nabla\phi_i|^2}
f_a^la_{l,i}\right)-\frac{\partial}{\partial
y}\left(\frac{\phi_{i,x}}{|\nabla\phi_i|^2}f_a^la_{l,i}\right)\Bigg],
\end{eqnarray}
with $\phi_{i,x}=\partial \phi_i/\partial x$. Anisotropy is
introduced using the orientation-dependent coefficient
\begin{equation}
a_i(\theta)=\left[\frac{1}{2}a_i(\theta_j)+\frac{1}{2}a_i(\theta_k)\right],\label{aij_aik}
\end{equation}
where $\theta$ is the angle between the interface normal and a reference crystal axis,
and
\begin{equation}
a_{l,i}=\frac{1}{2}\frac{\partial a_l(\theta_i)}{\partial
\theta_i}.\label{ali_deri}
\end{equation}
Since the orientation angle of the j-k binary interface can be
calculated by either
\begin{equation}
\sin\theta_j=-\partial_y\phi_j/\sqrt{(\partial_y\phi_j)^2+(\partial_x\phi_j)^2},
\end{equation}
or
\begin{equation}
\sin\theta_k=-\partial_y\phi_k/\sqrt{(\partial_y\phi_k)^2+(\partial_x\phi_k)^2},
\end{equation}
Eq. \ref{aij_aik} combines the contribution from both $\theta_j$ and
$\theta_k$ equally. To quantitatively incorporate an experimentally
relevant anisotropic surface energy, the coefficient
$a_i(\theta)$, which adjusts the height of the free-energy barrier between phases j and k,
can be directly related to $\gamma_{jk}(\theta)$ by the relation derived in appendix~\ref{growth_appendix_facet}
\begin{equation}
a_i(\theta)=B_0+B_1\frac{\gamma_{jk}(\theta)}{Wh}+B_2\left(\frac{\gamma_{jk}(\theta)}{Wh}\right)^2,
\end{equation}
with $B_0=-4.86349$, $B_1=-0.693313$ and $B_2=23.3564$.

\section{Sharp-interface limit of phase-field model}
\label{growth_si}

In this section, we carry out an asymptotic analysis to
relate the PF and sharp-interface models. In the present context, this analysis consists of deriving from the PF model the evolution equations for the solid-liquid and solid-vapor interfaces in the limit where the thickness of those spatially diffuse interfaces is small compared to the macroscopic scale of the system that is set by the NW radius, itself determined by the catalyst volume. The solid-vapor interface is assumed to have a vanishing mobility, consistent with the fact that surface diffusion is too slow to lead to a significant reconfiguration of this interface shape on the timescale of NW growth.
%Since all
%interfaces in PF models are represented by a diffuse region where PF
%parameters vary smoothly from one phase to another, it is necessary
%to show that the original physics described in sharp-interface model
%can be correctly recovered in PF model. Sharp-interface analysis is
%an analytical procedure carried on PF equations of motion. By
%expanding the motion of PF interface profile using a small
%parameter, which is typically the ratio between PF interface width
%and some other length scale in the system, this method is used to
%examine the convergence of PF interface motion to its
%sharp-interface counterpart if the width of the diffuse interface in
%PF model is small comparing with another intrinsic length scale in
%the problem of interest. Classic examples of sharp-interface
%analysis can be found in literatures of PF solidification
%models~\cite{Alain_quantitativePF_PRE}.

The present analysis is simpler than the thin interface limit of
solidification models insofar as interface motion is not coupled to a long range diffusion field such as temperature \cite{Alain_quantitativePF_PRE} or alloy concentration \cite{Alain_quantitativePF_PRL,Plapp}, i.e. the concentration of growth atom is assumed to be spatially uniform inside the catalyst. However, the analysis is made more complicated than the thin interface limit of solidification models by the introduction of a Lagrange multiplier
$\lambda_A$ to control the catalyst volume
(Eq.~\ref{func_vconstraint}). In order to first understand the effect of
this Lagrange multiplier in the simplest possible two-phase configuration, we
analyze in appendix~\ref{growth_si_appendix} the shape evolution of an isolated liquid droplet in a vapor phase, which can be described by a single order parameter $\phi$. We show that this evolution is governed in the sharp-interface limit by the equation
\begin{equation}
V=M\left(-\kappa\gamma+\gamma\frac{\int\kappa
ds}{\int ds }+\frac{\dot A}{\int ds}\right),\label{modified_Cahn_Allen}
\end{equation}
where $V$ is the normal interface velocity, $M$ is an interface mobility, $\kappa$ the is the interface curvature, $s$ is the arclength coordinate along the interface such that $\int ds$, evaluated along the closed interface contour surrounding the droplet, represents the total interface length, and $\dot A=dA(t)/dt$ where $A(t)$ is the droplet area. Eq. \ref{modified_Cahn_Allen} is simplest to interpret in the limit where the droplet area is constant ($\dot A=0$). In this case, it implies that an arbitrarily shaped droplet will relax to a circle while preserving the droplet area since $\dot A=\int Vds=0$. The motion involves both a local term $-M\kappa\gamma$, equivalent to motion by mean curvature, and an area-preserving nonlocal term $M\gamma\int\kappa ds/\int ds$. If $\dot A\ne 0$, Eq. \ref{modified_Cahn_Allen} implies that relaxation to a circle will occur concurrently with a change of droplet area since $\int Vds=\dot A$ in this case.

In this section, we extend the sharp-interface analysis to the multiphase-field model, which describes the more complex three-phase configuration where different regions of the liquid droplet surface are in contact with the vapor and solid phases.
In this case, Eq. \ref{modified_Cahn_Allen} takes on a more complex form that couples the evolution of the solid-liquid and liquid-vapor interfaces to $\dot A$, which only vanishes in the steady-state growth regime. Importantly, this evolution equation can be simplified and interpreted physically in the limit where the liquid-vapor interface mobility is much larger than the solid-liquid interface mobility. We show that, in this experimentally relevant limit where the droplet maintains a circular shape during growth, the Lagrange multiplier $\lambda_A$ reduces to the term corresponding to the Laplace pressure inside the droplet in the sharp-interface model, and the desired evolution equation for the solid-liquid interface dynamics can be obtained by a suitable choice of driving force (chemical potential difference $\Delta\mu$) in the PF model.

To carry out the sharp-interface analysis, the essential equations in
the PF model are reviewed here. The basic equations of motion
(Eqs.~\ref{iso_eqn1} and \ref{iso_eqn2}) are
\begin{eqnarray}
\nonumber\tau\frac{\partial \phi_1}{\partial
t}&=&K(\vec{\phi})\Bigg[W^2\nabla^2\phi_1-\frac{\partial
f_{p}}{\partial
\phi_1}\\
\nonumber&&+\Delta\tilde{\mu}u(\phi_1,\phi_3)+\lambda_A\frac{\partial
g_l}{\partial
\phi_1}-\frac{1}{3}\lambda_{\phi}\Bigg],\\
\label{asy_anal_eqn1}
\end{eqnarray}
\begin{eqnarray}
\nonumber\tau\frac{\partial \phi_2}{\partial
t}&=&K(\vec{\phi})\Bigg[W^2\nabla^2\phi_2-\frac{\partial
f_{p}}{\partial \phi_2}+\lambda_A\frac{\partial g_l}{\partial
\phi_2}-\frac{1}{3}\lambda_{\phi}\Bigg],\\
\label{asy_anal_eqn2}
\end{eqnarray}
with
\begin{equation}
\Delta\tilde{\mu}=\Delta\mu/(h\Omega_s),
\end{equation}
\begin{equation}
\lambda_{\phi}=\sum_{j=1}^3\bigg[W^2\nabla^2\phi_j-\frac{\partial
f_{p}}{\partial \phi_j}+\lambda_A\frac{\partial g_l}{\partial
\phi_j}\bigg],
\end{equation}
where we have used the expression of $f_p$ in
Eq.~\ref{pf_f_p}. The Lagrange multiplier $\lambda_A$ is
\begin{equation}
\lambda_A=\frac{I_1- I_2+\dot{A}\tau}{I_3-I_4},
\end{equation}
with the integrals $I_{1,2,3,4}$ defined in
Eqs.~\ref{int_I1}-\ref{int_I4}.

At the liquid-vapor binary interface, Eqs.~\ref{asy_anal_eqn1} and
\ref{asy_anal_eqn2} are reduced to
\begin{equation}
\tau\frac{\partial \phi_1}{\partial t}=0,
\end{equation}
\begin{equation}
\tau\frac{\partial \phi_2}{\partial
t}=K^{lv}\left(W^2\nabla^2\phi_2-\frac{1}{2}\frac{\partial
\tilde{f}_p^{lv}}{\partial
\phi_2}+\frac{1}{2}\lambda_A\frac{\partial \tilde{g_l}}{\partial
\phi_2}\right), \label{growth_app_eqn_mpf_motion_lv}
\end{equation}
where $\tilde{f}_p^{lv}$ and $K^{lv}$ are $f_p$ and $K(\vec{\phi})$
evaluated at the liquid-vapor interface ($\phi_1=0$,
$\phi_3=1-\phi_2$). The second term in the parenthesis on the right
of Eq.~\ref{growth_app_eqn_mpf_motion_lv} is obtained by summing the
$\partial f_p/\partial \phi_2$ term in Eq.~\ref{asy_anal_eqn2} with
all the $\partial f_p/\partial \phi_i$ terms in $\lambda_{\phi}$.
The third term is obtained in the same way.

At the solid-liquid interface, the PF equations of motion become
\begin{eqnarray}
\nonumber\tau\frac{\partial \phi_1}{\partial t}
&=&K^{sl}\bigg[W^2\nabla^2\phi_1-\frac{1}{2}\frac{\partial
\tilde{f}_p^{sl}}{\partial \phi_1}\\
&&-\frac{1}{2}\Delta\tilde{\mu}\frac{\partial \tilde{g_l}}{\partial
\phi_1}+\frac{1}{2}\lambda_A\frac{\partial \tilde{g_l}}{\partial
\phi_1}\bigg], \label{growth_app_eqn_mpf_motion_sl}
\end{eqnarray}
\begin{equation}
\tau\frac{\partial \phi_2}{\partial t}=0,
\end{equation}
where $\tilde{f}_p^{sl}$ and $K^{sl}$ are $f_p$ and $K(\vec{\phi})$
evaluated at the solid-liquid interface. For the $\Delta\tilde{\mu}$
term, $u(\phi_1,\phi_3)$ is reduced to $-\frac{1}{2}\frac{\partial
\tilde{g_l}}{\partial \phi_1}$ at the solid-liquid interface. Other
terms are obtained in the same way as in the liquid-vapor equation
(Eq.~\ref{growth_app_eqn_mpf_motion_lv}).

The sharp-interface asymptotics of these equations can be taken
directly from appendix~\ref{growth_si_appendix} by matching the PF
terms in Eqs.~\ref{growth_app_eqn_mpf_motion_lv} and
\ref{growth_app_eqn_mpf_motion_sl} to the single PF equation
(Eq.~\ref{growth_app_eqn_simplepf_2}) and taking the corresponding
asymptotics from the single PF result
(Eq.~\ref{growth_app_eqn_simplepf_exp4}). For the solid-liquid
interface, the sharp-interface equation is
\begin{equation}
v_{sl}=\frac{-\kappa_{sl}^0\gamma_{sl}^0+\frac{1}{2}(\lambda_A^1-\Delta\tilde{\mu})}{Q_{sl}},\label{si_v_sl_eqn_1}
\end{equation}
with
\begin{equation}
\gamma_{sl}^0=\int^{+\infty}_{-\infty}(\phi^0_{1,z_1})^2dz_1,
\end{equation}
\begin{equation}
Q_{sl}=\int^{+\infty}_{-\infty}\alpha^{sl}(\phi^0_{1,z_1})^2dz_1.
\end{equation}
Here, $\phi^0$ is the equilibrium PF boundary profile,
$\phi^0_{i,z}$ is the derivative of $\phi^0_i$ with respect to $z$,
$\alpha^{sl}=\tau D/(W^2K^{sl})$, $\kappa_{sl}^0$ is the scaled
solid-liquid interface curvature, $\lambda_A^1$ is the reduced
Lagrange multiplier and the direction of interface normal coordinate
$z_1$ points to the liquid phase (see
appendix~\ref{growth_si_appendix} for details). Similarly, the
liquid-vapor equation is
\begin{equation}
v_{lv}=\frac{\kappa_{lv}^0\gamma_{lv}^0+\frac{1}{2}\lambda_A^1}{Q_{lv}},\label{si_v_lv_eqn_1}
\end{equation}
with
\begin{equation}
\gamma_{lv}^0=\int^{+\infty}_{-\infty}(\phi^0_{2,z_2})^2dz_2,
\end{equation}
\begin{equation}
Q_{lv}=\int^{+\infty}_{-\infty}\alpha^{lv}(\phi^0_{2,z_2})^2dz_2.
\end{equation}
$\kappa_{lv}^0$ is the scaled liquid-vapor interface curvature,
$\alpha^{lv}=\tau D/(W^2K^{lv})$, and the direction of interface
normal coordinate $z_2$ points to the liquid phase. Since this
sharp-interface equation is based on the vapor phase $\phi_2$ which
has a negative curvature in the NW growth geometry, the minus sign
in front of the curvature term is removed to keep consistency with
the positive curvature convention in the ST sharp-interface model.

By replacing $v$ and $\kappa^0$ with their unscaled dimensional form
$v=V l_c/D$, $\kappa^0=l_c\kappa$ for both the solid-liquid and the
liquid-vapor interfaces, and using the dimensional interfacial
energy $\gamma$ (see appendix~\ref{growth_si_appendix}),
Eqs.~\ref{si_v_sl_eqn_1} and \ref{si_v_lv_eqn_1} become
\begin{equation}
V_{sl}=M_{sl}(-\kappa_{sl}\gamma_{sl}+\lambda_Ah-\Delta\mu\Omega_s^{-1}),\label{si_v_sl_eqn_2}
\end{equation}
\begin{equation}
V_{lv}=M_{lv}(\kappa_{lv}\gamma_{lv}+\lambda_Ah),\label{si_v_lv_eqn_2}
\end{equation}
with the solid-liquid interface curvature $\kappa_{sl}$, the
liquid-vapor curvature $\kappa_{lv}$ and interfacial energies
\begin{equation}
\gamma_{lv}=2Wh\int_{-\infty}^{+\infty}\left(\frac{\partial
\phi_2^0}{\partial z_2}\right)^2dz_2,
\end{equation}
\begin{equation}
\gamma_{sl}=2Wh\int_{-\infty}^{+\infty}\left(\frac{\partial
\phi_1^0}{\partial z_1}\right)^2dz_1.
\end{equation}
These two interfacial energy equations are reduced to
Eqs.~\ref{G_sl_eqn_sum} and \ref{G_lv_eqn_sum} with the well-known
equipartition relations $(\partial \phi_1^0/\partial
z_1)^2=\frac{1}{2}\tilde{f}_p^{sl}$ and $(\partial \phi_2^0/\partial
z_2)^2=\frac{1}{2}\tilde{f}_p^{lv}$, and the interface mobilities are given by
\begin{equation}
M_{lv}=\frac{W}{\tau
h}\frac{1}{2\int_{-\infty}^{+\infty}\left(\frac{\partial
\phi_2^0}{\partial z_2}\right)^2K_{lv}^{-1} dz_2},
\end{equation}
\begin{equation}
M_{sl}=\frac{W}{\tau
h}\frac{1}{2\int_{-\infty}^{+\infty}\left(\frac{\partial
\phi_1^0}{\partial z_1}\right)^2K_{sl}^{-1} dz_1}.\label{mob_sl}
\end{equation}
The result in Eq.~\ref{M_sl_eqn_sum} is obtained by substituting in the
expression of $K$ (Eq.~\ref{growth_mob_K}) at a given binary
interface into Eq.~\ref{mob_sl}. The mobility ratio in
Eq.~\ref{growth_mob_ratio} is also derived from here using the
equipartition relations.

The sharp-interface counterpart of the Lagrange multiplier
$\lambda_A$ is derived following the procedure demonstrated in
appendix~\ref{growth_si_appendix} (Eqs.~\ref{app_lag_derive_eq1} -
\ref{growth_app_eqn_simplepf_lag_res}). The catalyst volume
condition in Eq.~\ref{growth_app_b4} can be separated into two parts
which cover the solid-liquid and liquid-vapor interfaces separately
as the following
\begin{eqnarray}
&&\dot{A}=\int\sum_{i=1}^2\left(\frac{\partial \tilde{g_l}}{\partial
\phi_i}\frac{\partial \phi_i}{\partial
t}\right)dv\\
\nonumber&&=\int_{sl}-V_{sl}\frac{\partial\tilde{g_l}}{\partial
\phi_1}\frac{\partial \phi_1}{\partial
z_1}dsdz_1+\int_{lv}-V_{lv}\frac{\partial\tilde{g_l}}{\partial
\phi_2}\frac{\partial \phi_2}{\partial
z_2}dsdz_2.\label{app_volume_change_si}
\end{eqnarray}
This is the multiphase-field equivalent of the single phase-field PF model in
Eq.~\ref{app_volume_change}. Replacing the interface velocities in
Eq.~\ref{app_volume_change_si} with Eqs.~\ref{si_v_sl_eqn_2} and
\ref{si_v_lv_eqn_2} and solving for the Lagrange multiplier
$\lambda_A$ gives
\begin{eqnarray}
\nonumber
\lambda_Ah&=&\frac{\dot{A}-\gamma_{lv}\kappa_{lv}M_{lv}S_{lv}+(\gamma_{sl}\kappa_{sl}+\Delta\mu\Omega_s^{-1})M_{sl}S_{sl}}{M_{lv}S_{lv}+M_{sl}S_{sl}},\label{lag_mpf}\\
\end{eqnarray}
which is the multiphase-field version of
Eq.~\ref{growth_app_eqn_simplepf_lag_res} with interface lengths
$S_{lv}$ and $S_{sl}$ defined by
\begin{equation}
S_{lv}=\int_{lv}ds,
\end{equation}
\begin{equation}
S_{sl}=\int_{sl}ds.
\end{equation}
In addition, the solid-liquid and liquid-vapor interface curvatures are
assumed to be constant in the derivation of Eq.~\ref{lag_mpf}.

Inserting Eq.~\ref{lag_mpf} into Eqs.~\ref{si_v_sl_eqn_2} and
\ref{si_v_lv_eqn_2} gives the sharp-interface equations of motion
for the solid-liquid and liquid-vapor interfaces
\begin{eqnarray}
\nonumber
V_{sl}&=&-M_{sl}\kappa_{sl}\gamma_{sl}-M_{sl}\Delta\mu\Omega_s^{-1}\\
\nonumber&&+M_{sl}\frac{-\kappa_{lv}\gamma_{lv}M_{lv}S_{lv}+(\kappa_{sl}\gamma_{sl}+\Delta\mu\Omega_s^{-1})M_{sl}S_{sl}+\dot{A}}{M_{lv}S_{lv}+M_{sl}S_{sl}}\\
&=&\frac{-\Delta\mu\Omega_s^{-1}-\kappa_{sl}\gamma_{sl}-\kappa_{lv}\gamma_{lv}+\frac{\dot{A}}{M_{lv}S_{lv}}}{\left(\frac{1}{M_{sl}}+\frac{S_{sl}}{S_{lv}}\frac{1}{M_{lv}}\right)}.
\end{eqnarray}
\begin{eqnarray}
\nonumber V_{lv}&=&M_{lv}\kappa_{lv}\gamma_{lv}\\
\nonumber&&+M_{lv}\frac{-\kappa_{lv}\gamma_{lv}M_{lv}S_{lv}+(\kappa_{sl}\gamma_{sl}+\Delta\mu\Omega_s^{-1})M_{sl}S_{sl}+\dot{A}}{M_{lv}S_{lv}+M_{sl}S_{sl}}\\
&=&\frac{\Delta\mu\Omega_s^{-1}+\kappa_{sl}\gamma_{sl}+\kappa_{lv}\gamma_{lv}+\frac{\dot{A}}{M_{sl}S_{sl}}}{\left(\frac{S_{lv}}{S_{sl}}\frac{1}{M_{sl}}+\frac{1}{M_{lv}}\right)}.
\end{eqnarray}

These expressions can be further simplified in the limit
$M_{lv}\gg M_{sl}$ where the liquid droplet relaxes quasi-instantaneously to an equilibrium shape during growth.
Neglecting the volume change
contribution $\dot{A}$, which is typically small and vanishes in the steady-state growth regime,
Eq.~\ref{lag_mpf} reduces in this limit to
\begin{equation}
\lambda_Ah\approx-\kappa_{lv}\gamma_{lv},\label{lagrange_si_reduced}
\end{equation}
which corresponds to the Laplace pressure of the droplet. The
sharp-interface limit of the solid-liquid interface motion becomes
\begin{equation}
v_n=V_{sl}=M_{sl}(-\kappa_{sl}\gamma_{sl}-\frac{\Delta\mu}{\Omega_s}-\kappa_{lv}\gamma_{lv}).
\label{growth_app_eqn_mpf_motion_solid_limit_final}
\end{equation}
We note that, with the volume change factor $\dot{A}$ included, the expression for the Lagrange
multiplier contains an additional contribution
\begin{equation}
\lambda_Ah\approx\frac{\dot{A}}{M_{lv}S_{lv}}-\kappa_{lv}\gamma_{lv}\label{lagrange_si_reduced_with_dotA}.
\end{equation}
However, the same solid-liquid interface equation of motion
(Eq.~\ref{growth_app_eqn_mpf_motion_solid_limit_final}) is obtained
since $\dot{A}$ represents a high order contribution in the limit where $M_{sl}/M_{lv}\ll 1$. As a result, the rapid
volume change mediated by the motion of the liquid-vapor interface
has a negligibly small effect on the NW growth rate in this limit, as physically desired.

Finally, Eq.~\ref{growth_app_eqn_mpf_motion_solid_limit_final} can be readily seen to have the same form as the
equation for the normal velocity of the solid-liquid interface in the sharp-interface model, which is obtained by combining Eqs.~\ref{sl_growth_law}-\ref{sl_mus}
\begin{equation}
v_n=M_{sl}\left[\frac{\beta(c_l-c_0)}{\Omega_s}+\kappa_{lv}\gamma_{lv}\frac{\Omega_l}{\Omega_s}-\kappa_{sl}\gamma_{sl}-\kappa_{lv}\gamma_{lv}\right],
\label{growth_app_eqn_mpf_motion_sharpinterface}
\end{equation}
if $\Delta\mu$ is chosen in the PF model as
\begin{eqnarray}
\nonumber\Delta\mu&=&-[\beta(c_l-c_0)+\kappa_{lv}\gamma_{lv}\Omega_l]\\
&=&-[\beta(c_l-c_0)-\lambda_Ah\Omega_l],\label{growth_mu_tersoff}
\end{eqnarray}
where Eq.~\ref{lagrange_si_reduced} is used in the second equality.
By the same argument given above, this $\Delta\mu$ expression remains valid when $\dot{A}\ne 0$ since
the slow catalyst volume change has a negligible effect on the solid-liquid interface dynamics in
the  rapid droplet-shape relaxation limit $M_{sl}/M_{lv}\ll 1$.

%The ST model also has a dynamics for the number of Si atoms $N_g$,
%and the over-saturation $c_l-c_0$ is tuned during the growth through
%this $N_g$ dynamics. The same atom number dynamics can be easily
%re-written in terms of phase-field as shown in Eq.~\ref{pf_dndt}.
%However, the purpose of that part of the model is to give the
%concentration related driving force for NW growth which will be
%translated into a term in the chemical potential difference
%$\Delta\mu$ in the PF model, therefore it has essentially no impact
%on the sharp-interface limit of the PF dynamics.

\section{Numerical examples and comparison with sharp-interface theory}
\label{num_results}
In this section, we discuss the numerical
implementation of the phase-field model. We then present results of
simulations that illustrate the ability of the model to reproduce
basic features of NW growth. We consider first the simpler case of
an isotropic solid-liquid interface and then consider the more
realistic case of a faceted solid-liquid interface. The quantitative
validity of the approach is tested by comparisons with
sharp-interface theory for the NW growth shape and velocity.

\subsection{Numerical implementation}

%The PF equations (Eqs. \ref{iso_eqn1}, \ref{iso_eqn2}, and \ref{iso_eqn3}), the
%concentration evolution equation (Eqs. \ref{pf_dndt} and \ref{pf_mu}) and the
%catalyst volume evolution equation (Eq. \ref{pf_dvdt}) are first written in dimensionless form by introducing
%the dimensionless time
%$\bar{t}=t/\tau$ and dimensionless length $\bar{x}=x/W$ (and the
%corresponding dimensionless derivative operator
%$\bar{\nabla}=W\nabla$, volume element $d\bar{v}=dv/W^2$ and
%catalyst size $\bar{A}=A/W^2$).

The PF equations are first written in dimensionless form by introducing
the dimensionless time
$\bar{t}=t/\tau$ and dimensionless length $\bar{x}=x/W$ as well as the
corresponding dimensionless derivative operator
$\bar{\nabla}=W\nabla$, volume element $d\bar{v}=dv/W^2$, and
catalyst size $\bar{A}=A/W^2$. The functional derivatives that include the driving force for crystallization become
\begin{eqnarray}
\nonumber\frac{1}{h}\frac{\hat{\delta}F}{\hat{\delta}\phi_1}&=&-\bigg[\bar{\nabla}^2\phi_1-2\phi_1(1-\phi_1)(1-2\phi_1)\\
&&-\sum_{i=1}^3\frac{\partial (a_if_a^i+b_if_b)}{\partial
\phi_1}+\Delta\tilde{\mu}u(\phi_1,\phi_3)\bigg],\label{functional_derivative_1_numerics}
\end{eqnarray}
\begin{eqnarray}
\nonumber\frac{1}{h}\frac{\hat{\delta}F}{\hat{\delta}\phi_2}&=&-\bigg[\bar{\nabla}^2\phi_2-2\phi_2(1-\phi_2)(1-2\phi_2)\\
&&-\sum_{i=1}^3\frac{\partial (a_if_a^i+b_if_b)}{\partial
\phi_2}\bigg],\label{functional_derivative_2_numerics}
\end{eqnarray}
\begin{eqnarray}
\nonumber\frac{1}{h}\frac{\hat{\delta}F}{\hat{\delta}\phi_3}&=&-\bigg[\bar{\nabla}^2\phi_3-2\phi_3(1-\phi_3)(1-2\phi_3)\\
&&-\sum_{i=1}^3\frac{\partial (a_if_a^i+b_if_b)}{\partial
\phi_3}-\Delta\tilde{\mu}u(\phi_1,\phi_3)\bigg],\label{functional_derivative_3_numerics}
\end{eqnarray}
with
\begin{equation}
\Delta\tilde{\mu}=-\left[\frac{\beta}{h\Omega_s}(c_l-c_0)-\lambda_A\frac{\Omega_l}{\Omega_s}\right].
\end{equation}
Using
Eqs. \ref{functional_derivative_1_numerics}-\ref{functional_derivative_3_numerics},
the Lagrange multiplier $\lambda_A$ can be evaluated as
\begin{equation}
\lambda_A=\frac{I_1- I_2+\frac{\partial
\bar{A}}{\partial\bar{t}}}{I_3-I_4},\label{lag_sum_I_num}
\end{equation}
where
\begin{equation}
I_1=\int Kh^{-1}\sum_{i=1}^2 \frac{\hat{\delta} F}{\hat{\delta}
\phi_i}\frac{\partial \tilde{g_l}}{\partial \phi_i}d\bar{v},
\end{equation}
\begin{equation}
I_2=\frac{1}{3}\int Kh^{-1}\sum_{i=1}^2 \frac{\partial
\tilde{g_l}}{\partial \phi_i}\sum_{j=1}^3\frac{\hat{\delta}
F}{\hat{\delta} \phi_j}d\bar{v},
\end{equation}
\begin{equation}
I_3=\int K\sum_{i=1}^2\left(\frac{\partial
\tilde{g_l}}{\partial\phi_i}\right)^2 d\bar{v},
\end{equation}
\begin{equation}
I_4=\frac{1}{3}\int K\left(\sum_{i=1}^2 \frac{\partial
\tilde{g_l}}{\partial \phi_i}\right)^2d\bar{v}.
\end{equation}
The phase-field evolution equations are
\begin{equation}
\frac{\partial\phi_j}{\partial\bar{t}}=-\frac{K(\vec{\phi})}{h}\left(\frac{\hat{\delta}\tilde{F}}{\hat{\delta}\phi_j}-\frac{1}{3}\sum_{i=1}^3\frac{\hat{\delta}\tilde{F}}{\hat{\delta}\phi_i}\right),\label{dphi_dt_num}
\end{equation}
for $j=$1 and 2 with
\begin{equation}
\frac{1}{h}\frac{\hat{\delta}\tilde{F}}{\hat{\delta}\phi_i}=\frac{1}{h}\frac{\hat{\delta}F}{\hat{\delta}\phi_i}-\lambda_A\frac{\partial
g_l}{\partial \phi_i},
\end{equation}
and
\begin{equation}
\phi_3=1-\phi_1-\phi_2.
\end{equation}
The evolution equations for the catalyst concentration $c_l=(N_c+N_g)/N_c$ and catalyst volume (area in 2D) are determined by
\begin{equation}
\frac{dN_g}{d\bar{t}}=\frac{JW\tau}{\bar{\eta}}\int \phi_2\phi_3
d\bar{v}-\frac{W^2}{\Omega_s}\int \frac{\partial g_s}{\partial
\bar{t}}d\bar{v},\label{pf_dndt_num}
\end{equation}
which determines the evolution of $c_l$ at fixed number of catalyst atoms $N_c$, and
\begin{equation}
\frac{d\bar{A}}{d\bar{t}}=\frac{\Omega_l}{W^2}\frac{dN_g}{d\bar{t}},\label{pf_dvdt_num}
\end{equation}
respectively, where $\bar{\eta}=\eta/W$ and in 2D
the incorporation flux $J$ has the unit of atoms per length per time.

To model an anisotropic surface energy, the $\frac{\partial (a_if_a^i+b_if_b)}{\partial \phi_j}$
term in Eqs. \ref{functional_derivative_1_numerics}-\ref{functional_derivative_3_numerics}
is replaced by the form defined by Eq. \ref{facet_fd1}
\begin{align}
\nonumber\sum_{i=1}^3&\frac{\partial (a_if_a^i+b_if_b)}{\partial
\phi_j}\rightarrow&\\
\nonumber&\sum_{i=1}^3\Bigg[a_i\frac{\partial f_a^i}{\partial
\phi_j}+b_i\frac{\partial f_b}{\partial
\phi_j}+\frac{\partial}{\partial
x}\left(\frac{\phi_{j,y}}{|\nabla\phi_j|^2}
f_a^ia_{i,j}\right)&\\&-\frac{\partial}{\partial
y}\left(\frac{\phi_{j,x}}{|\nabla\phi_j|^2}f_a^ia_{i,j}\right)\Bigg].&\label{facet_fd_num}
\end{align}
The constant barrier parameter $a_i$ in the isotropic model becomes
orientation dependent $a_i(\theta)$ in the anisotropic model. The
orientation dependent barrier parameter $a_i$ and its derivative
$a_{i,j}$ are defined in Eqs.~\ref{aij_aik} and \ref{ali_deri}. To
model a given anisotropic solid-liquid interfacial energy
$\gamma(\theta)$, $a_i(\theta)$ needs to be calculated numerically
using Eq.~\ref{a_gamma_inversion}.

The phase-field evolution equations (Eq. \ref{dphi_dt_num}) are
stepped forward in time with an explicit Euler scheme with centered
finite difference approximations of the spatial derivates. The
evolution equations for the concentration (Eq. \ref{pf_dndt_num})
and catalyst volume (Eq. \ref{pf_dvdt_num}) are also stepped forward
in time with an explicit Euler scheme with parameters $c_0=0.45$,
$\beta\Omega_s^{-1}h^{-1}=2.0$ and $\Omega_s/W^2=\Omega_l/W^2=1.0$
unless explicitly specified otherwise. Space and time
discretizations are chosen to be $\Delta x/W=\Delta y/W=0.4$ and
$\Delta t/\tau=0.001$. To keep the numerics tractable, we use a
mobility ratio $M_{lv}/M_{sl}\approx 20$, corresponding to
$\alpha\approx 110$ in the expression of $K(\vec{\phi})$
(Eq.~\ref{K_mob_eqn_sum}), which is sufficiently large for the
liquid-vapor interface to relax to a circular equilibrium shape on
the characteristic time scale of NW growth. The liquid-vapor surface
energy is isotropic and given by $a_1=0$. The scaling factor in
Eq.~\ref{pf_dndt_num} $\bar{\eta}=0.71$. The other two surface
energy parameters ($a_2$ and $a_3$) are computed using Eq.
\ref{a_gamma_inversion} to match desired ratios of interfacial
free-energies. The parameter $b_i$ in the potential function $f_p$
is set to 80 to reduce the triple junction size. To increase
performance, only grid points near the liquid phase are computed.

\subsection{From droplet to nanowire}

\begin{figure}[!h]
  \includegraphics[width=80mm]{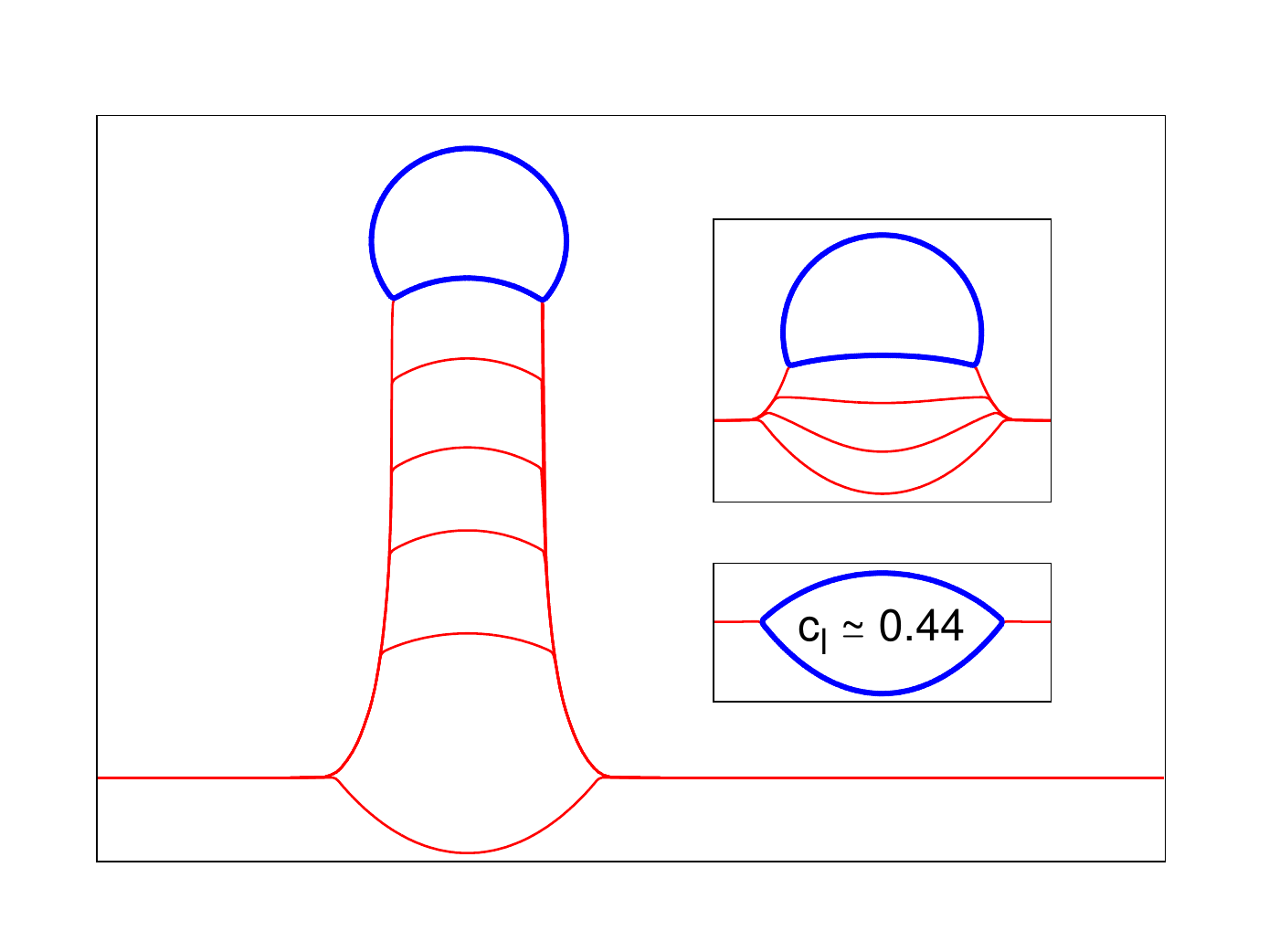}\\
  \caption{PF simulation with isotropic interfaces illustrating the evolution from droplet to NW. The solid-liquid and solid-vapor interfaces are shown as red lines at different times and the solid-liquid and liquid-vapor interfaces bounding the catalyst droplet are only shown together for clarity as a thicker blue line at the latest time. The top right inset shows interfaces more closely spaced in time during the initial growth stage. Simulation parameters are $A_0/W^2=313$, $JW\tau=0.007$. The lower right inset shows the equilibrium configuration on the substrate before growth ($J=0$). }\label{NW_grow_3step}
\end{figure}

Here we first demonstrate some basic features of this PF NW growth model
based on isotropic interfacial free-energies. For the silicon-gold system, we
use the values $\gamma_{sv}=1.2$ Jm$^{-2}$,
$\gamma_{sl}=0.8$ Jm$^{-2}$ and $\gamma_{lv}=1.0$ Jm$^{-2}$ which are
similar to those given in previous studies~\cite{Roper_voorhees_2007}.
The initial configuration is a substrate-vapor system. By seeding a
catalyst droplet of a specified volume on the substrate with
$c_l=c_0$, without Si incorporation at the liquid-vapor interface,
the catalyst relaxes to an equilibrium shape given by Young's condition. Once the
flux of Si atom is switched on at the liquid-vapor interface, the NW
grows vertically as demonstrated in Fig. \ref{NW_grow_3step}.

\begin{figure}[!h]
  \includegraphics[width=80mm]{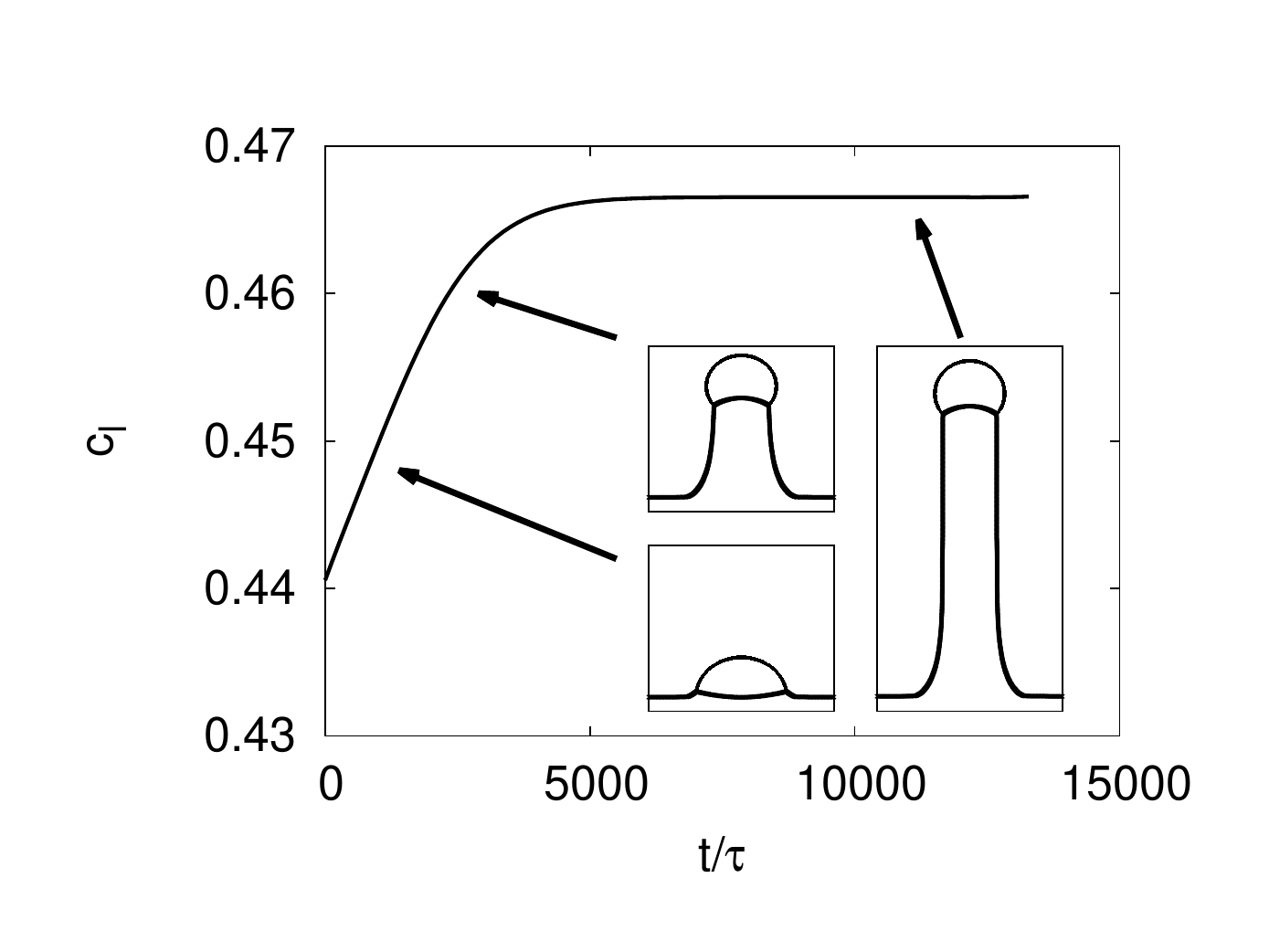}\\
  \caption{Droplet concentration versus time scaled by the phase-field relaxation time $\tau$
  during NW growth for the same parameters as Fig.~\ref{NW_grow_3step}.
Insets show morphologies during initial growth, tapering, and stead-state growth.}\label{NW_grow_ct}
\end{figure}

\begin{figure}[!h]
  \includegraphics[width=80mm]{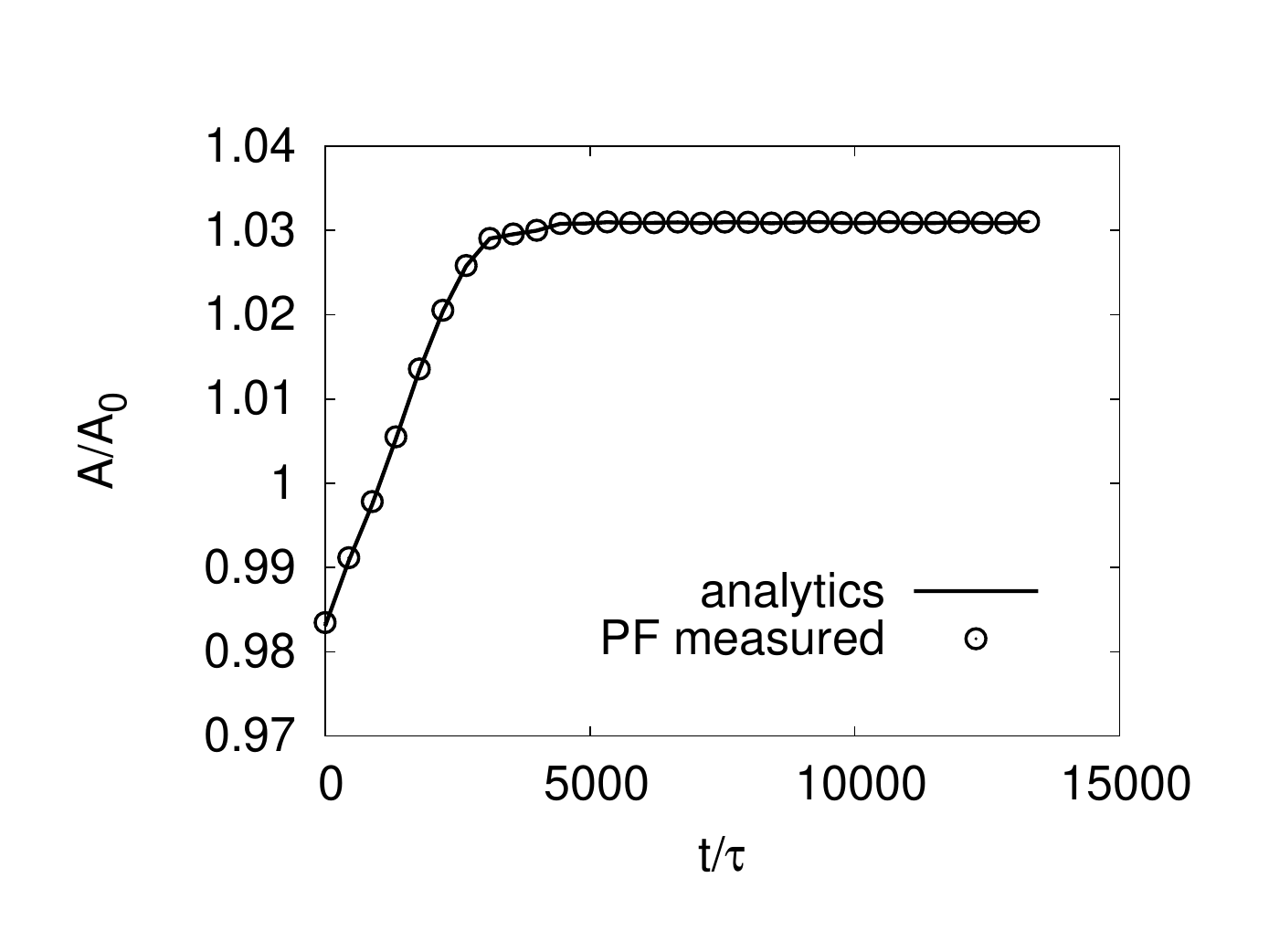}\\
  \caption{Catalyst area $A$ scaled by the area of an equilibrium droplet $A_0$ before growth versus dimensionless time $t/\tau$ with $A$ analytically predicted by
 Eq.~\ref{v_change_analytic} with $c_l$ from the PF simulation of Fig.~\ref{NW_grow_ct} (solid line) and with $A$ computed from the same simulation using $\int g_l(\vec \phi) dv$ that defines the catalyst area in Eq.~\ref{func_vconstraint} (open circles).
Simulation parameters are the same as in Fig.~\ref{NW_grow_3step} and Fig.~\ref{NW_grow_ct}.}\label{NW_grow_vt}
\end{figure}

The catalyst concentration and volume
during growth are shown in Fig. \ref{NW_grow_ct} and
Fig. \ref{NW_grow_vt}, respectively. The catalyst is under-saturated at the
start of growth due to the Gibbs-Thompson effect associated with the curvature of the liquid-vapor
surface and becomes over-saturated as growth atoms become incorporated in the catalyst droplet. Finally,
the growth velocity and droplet concentration reach constant values during steady-state growth. For
a given catalyst size, the volume evolution described by
Eq. \ref{pf_dvdt} is also accurately reproduced as shown in
Fig. \ref{NW_grow_vt}.


\subsection{Nanowire radius for steady-state growth}

\begin{figure}[!h]
\center
  \includegraphics[width=60mm]{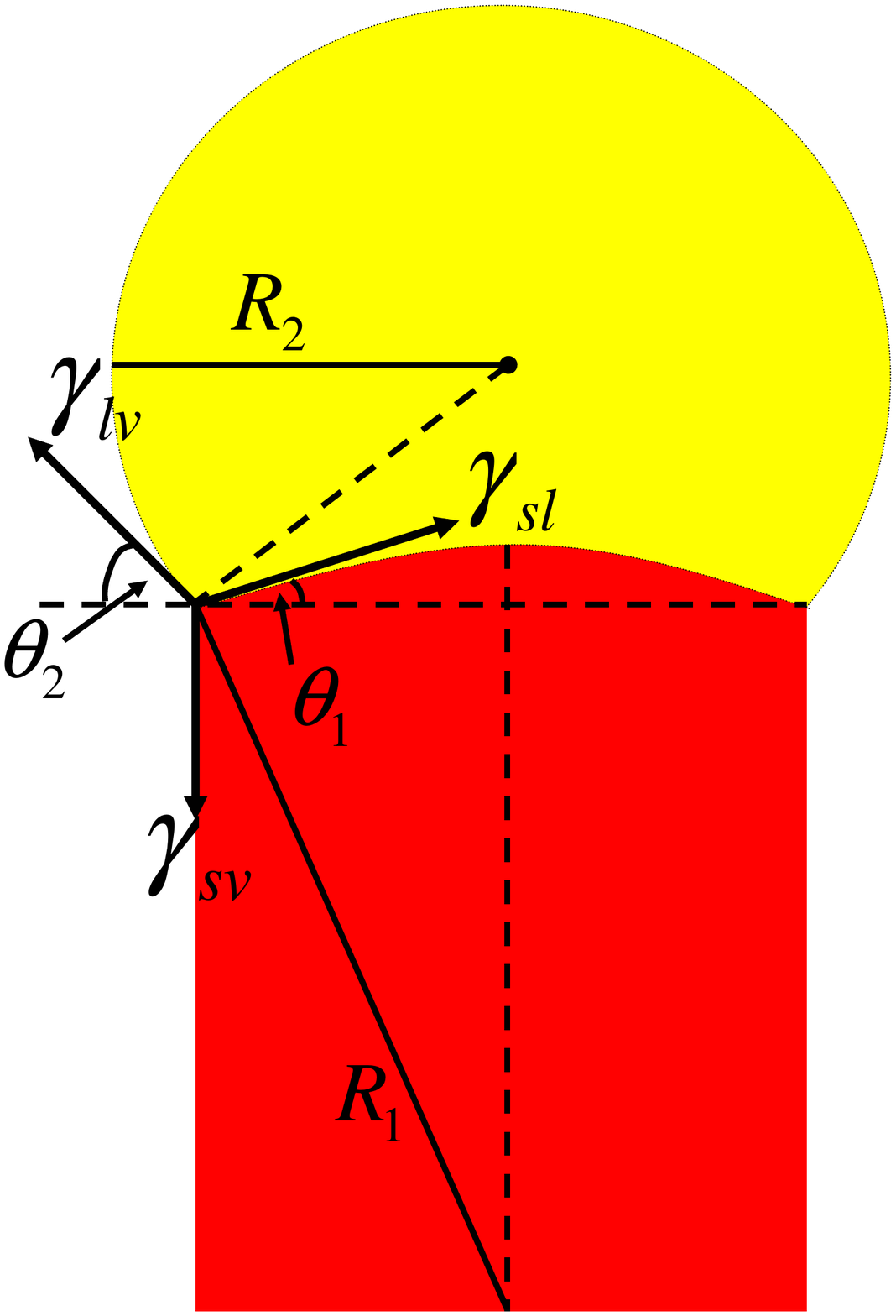}\\
  \caption{Definition of geometrical parameters used to characterize steady-state NW growth.
 The NW (red region) and droplet catalyst (yellow region) are shown together with the radius of curvature of the solid-liquid (liquid-vapor) interface $R_1$ ($R_2$), the excess free-energies for the solid-liquid ($\gamma_{sl}$), liquid-vapor ($\gamma_{lv}$), and solid-vapor ($\gamma_{sv}$) interfaces, and the corresponding dihedral angles determined by Young's condition at the three-phase junction, which are defined here as the angle between a horizontal line and the solid-liquid interface ($\theta_1$) and the liquid-vapor interface ($\theta_2$).} \label{wire_angles} \center
\end{figure}

We now compare the steady-state NW growth shape to the prediction of sharp-interface theory.
The steady-state NW radius predicted by sharp-interface theory, denoted here as $R_{si}$,  is determined by the three interfacial free-energies together with the
size of the catalyst (as shown in Fig. \ref{wire_angles}). For isotropic interfaces, the
Young-Herring condition reduces to Young's condition at the
triple-phase junction. For a vertical sidewall, the projection of the capillary forces on the vertical and horizontal directions yield the relations
\begin{equation}
\gamma_{sv}=\gamma_{sl}\sin\theta_1+\gamma_{lv}\sin\theta_2,\label{si_gamma_balance_y}
\end{equation}
and
\begin{equation}
\gamma_{sl}\cos\theta_1=\gamma_{lv}\cos\theta_2,
\end{equation}
respectively.

\begin{figure}[!h]
\center
  \includegraphics[width=80mm]{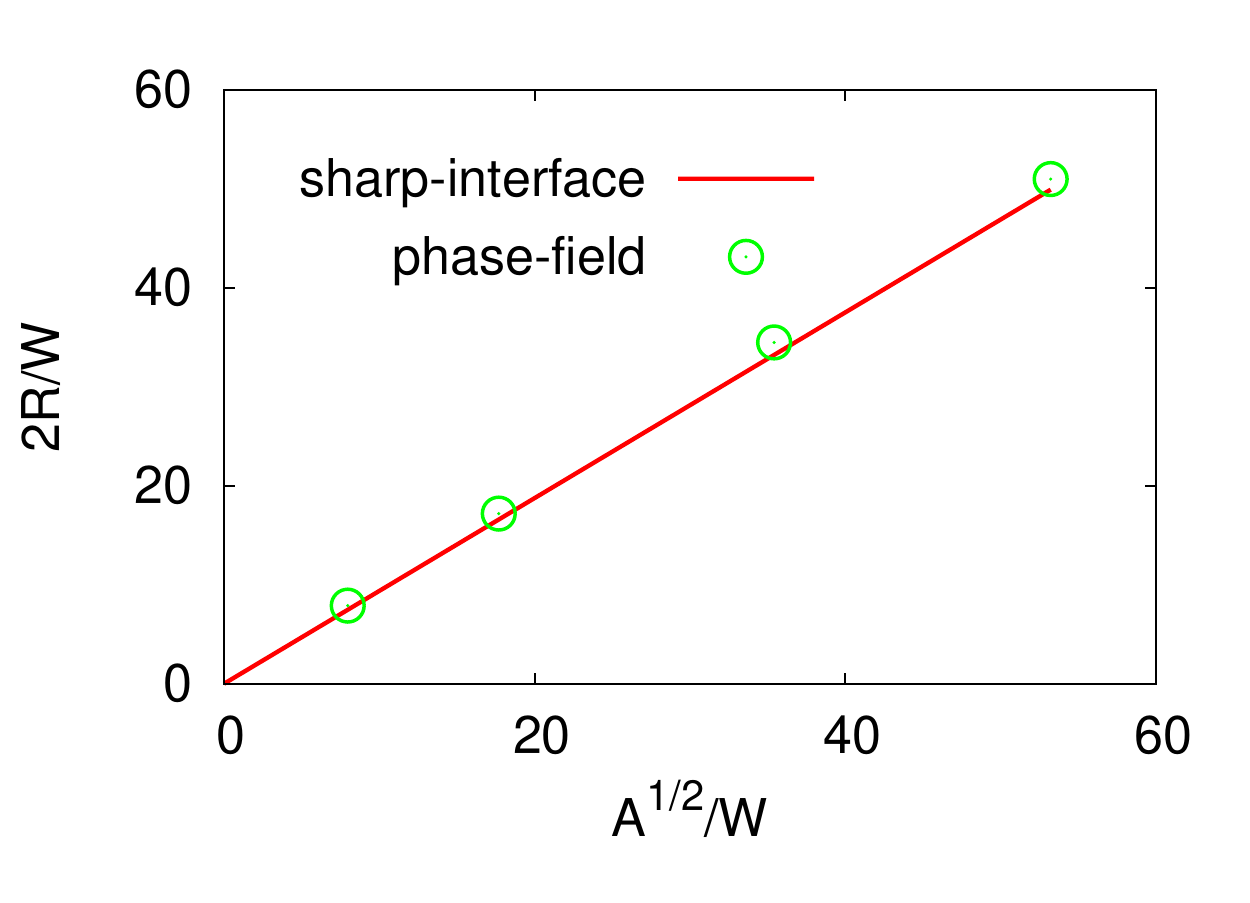}\\
  \caption{NW diameter $2R$ normalized by the interface thickness $W$ versus dimensionless catalyst size defined as $A^{1/2}/W$ where $A$ is the steady-state value of the catalyst area. The analytical prediction of the sharp-interface model (Eq. \ref{r_a_linear} and red line) is compared to the results of PF simulations (open green circles).
  A constant flux $JW\tau=0.0014$ is used and other parameters are the same as in Fig. \ref{NW_grow_3step}.}
  \label{dia_profile} \center
\end{figure}

The radius of curvature of the solid-liquid ($R_1$) and the liquid-vapor
($R_2$) interfaces are related to the NW radius $R_{si}$ by
\begin{equation}
R_1=R_{si}/\sin\theta_1,
R_2=R_{si}/\sin\theta_2,\label{si_R_relations}
\end{equation}
from which we obtain the expression for the catalyst area
\begin{equation}
A=\pi
R_2^2-\theta_2R_2^2+R_2^2\sin\theta_2\cos\theta_2-\theta_1R_1^2+R_1^2\sin\theta_1\cos\theta_1.
\end{equation}
This expression can be further reduced to
\begin{equation}
A^{1/2}=R_{si}\left(\frac{\pi-\theta_2}{\sin^2\theta_2}-\frac{\theta_1}{\sin^2\theta_1}+\cot\theta_1+\cot\theta_2\right)^{1/2},\label{r_a_linear}
\end{equation}
which predicts that the sharp-interface NW radius $R_{si}$ is proportional to
the square root of the catalyst area.
As shown in Fig. \ref{dia_profile}, this prediction is in very good quantitative agreement  with PF simulations where the catalyst area was varied over a very broad range.

\subsection{Steady-state nanowire growth rate}

%In the previous section, shape convergence of our PF model to the
%sharp-interface limit was demonstrated by a comparison of the NW
%radius from the two approaches.
The dependence of the NW growth rate on radius has been extensively
studied experimentally~\cite{Schmidt_dia,Lund_diameter_2007,Dubrovskii2008,dia_independent_growth}.
Both size-dependent and size-independent growth rates have been reported
in different experimental settings.
In this section, we examine the
convergence of the NW growth rate in the PF model to its sharp-interface
asymptotics.
We consider two physically distinct growth regimes.
The first is the one where
growth is limited by the solid-liquid interface kinetics and the
chemical potential of growth atoms can be assumed to be equal in the
liquid and vapor and constant in time, i.e. those two phases
equilibrate quickly on the time scale where the solid adds one
additional layer of atoms. In this regime, the growth rate depends on catalyst size.
The second is the one considered by ST where the NW growth rate is
limited by the incorporation rate of growth atoms at the droplet
surface. Since the total number of incorporated Si atoms is
proportional to the droplet surface area which is geometrically
related to NW radius, the growth rate in this case becomes
size-independent and is controlled only by the droplet
incorporation rate.

\begin{figure}[!h]
\center
  \includegraphics[width=80mm]{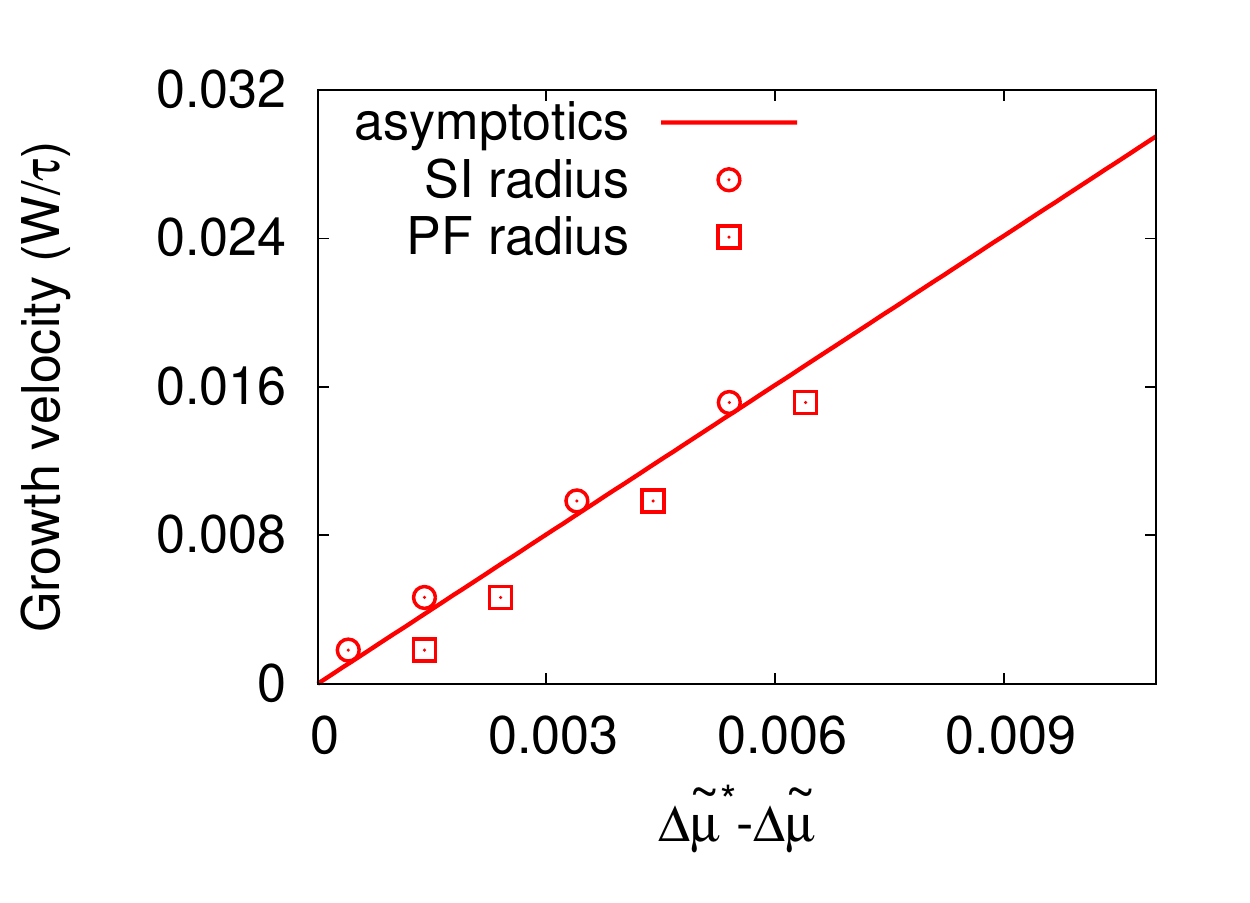}\\
  \caption{Comparison of steady-state NW growth rate versus driving force in PF simulations
  and predicted by sharp-interface theory for an initial catalyst area $A_0=1254\,W^2$.
  By calculating the critical driving force using Eq. \ref{mu_star},
  the growth rate (Eq. \ref{growth_app_eqn_mpf_motion_solid_limit_final}) is well-reproduced in PF
  simulations as shown by sharp-interface (SI) radius data points. This comparison is
  very sensitive to the NW radius. The growth rate computed by replacing $R_{si}$ in
  Eq.~\ref{mu_star} by the slightly different PF radius shown in
  Fig. \ref{dia_profile} (PF radius points) produces
  a noticeable shift of the growth threshold.}
  \label{v_mu_2} \center
\end{figure}

For the first interface-kinetics-dominated regime, the growth rate is determined
by Eq. \ref{growth_app_eqn_mpf_motion_solid_limit_final} derived in the sharp-interface analysis of our PF model
(section. \ref{growth_si}) with a constant $\Delta\mu$.
%As seen in the sharp-interface expansion of our PF model
%in section. \ref{growth_si}, the NW growth rate is given by a simple
%formula Eq.\ref{growth_app_eqn_mpf_motion_solid_limit_final}.
This equation implies that the NW growth rate vanishes at a critical driving force
$\Delta\mu^*=-(\kappa_{sl}\gamma_{sl}+\kappa_{lv}\gamma_{lv})\Omega_s$. Using Eq. \ref{si_R_relations} for the radii of curvature and Eq. \ref{si_gamma_balance_y}, this critical driving force
can be simplified to
\begin{equation}
\Delta\mu^*=-\gamma_{sv}\Omega_s/R_{si}.\label{mu_star}
\end{equation}
With a given $\Delta\mu$, the growth rate is then proportional to
$\Delta\mu^*-\Delta\mu$. Such a growth law can be easily tested in
our PF model by carrying out simulations that run long enough to
reach the steady-state growth regime with $\Delta\mu$ held constant.
Since the concentration dynamics described in Eqs.~\ref{pf_dndt},
\ref{pf_mu} and \ref{pf_dvdt} is not needed in these constant
$\Delta\mu$ simulations, catalyst volumes are set to $A=A_0$. PF
simulation results shown in Fig. \ref{v_mu_2} confirm the predicted
linear dependence of the NW growth rate on $\Delta\mu^*-\Delta\mu$.

\begin{figure}[!h]
  \includegraphics[width=80mm]{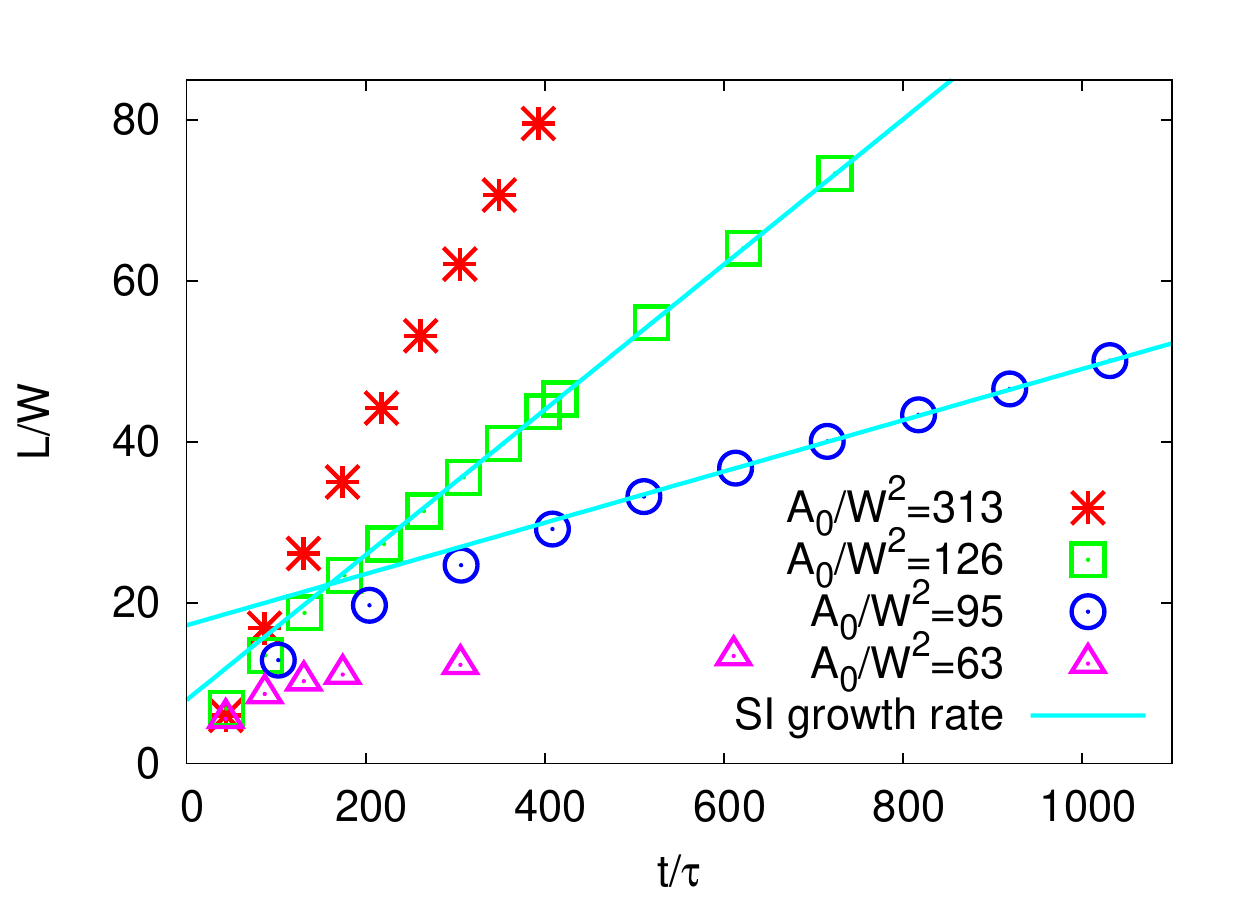}\\
  \caption{Scaled NW growth length as a function of scaled time for different catalyst
sizes in the interface-kinetics-dominated regime where the droplet
chemical potential is held constant (here $\Delta\tilde{\mu}=-0.14$) instead of being determined by Eqs.~\ref{pf_dndt}, \ref{pf_mu}, and
 \ref{pf_dvdt}. The results of PF simulations (symbols) agree well with the prediction of sharp-interface (SI) theory based on
 Eq.~\ref{growth_app_eqn_mpf_motion_solid_limit_final} (lines).}\label{l_mu}
\end{figure}

NWs of different size are then grown using this model in the
interface-kinetics-dominated regime by seeding the growth with
different catalyst sizes. Since the critical driving force for NW
growth depends on the solid-liquid and liquid-vapor curvatures, the
growth rate becomes size-dependent as shown in Fig.~\ref{l_mu}.

\begin{figure}[!h]
\center
  \includegraphics[width=80mm]{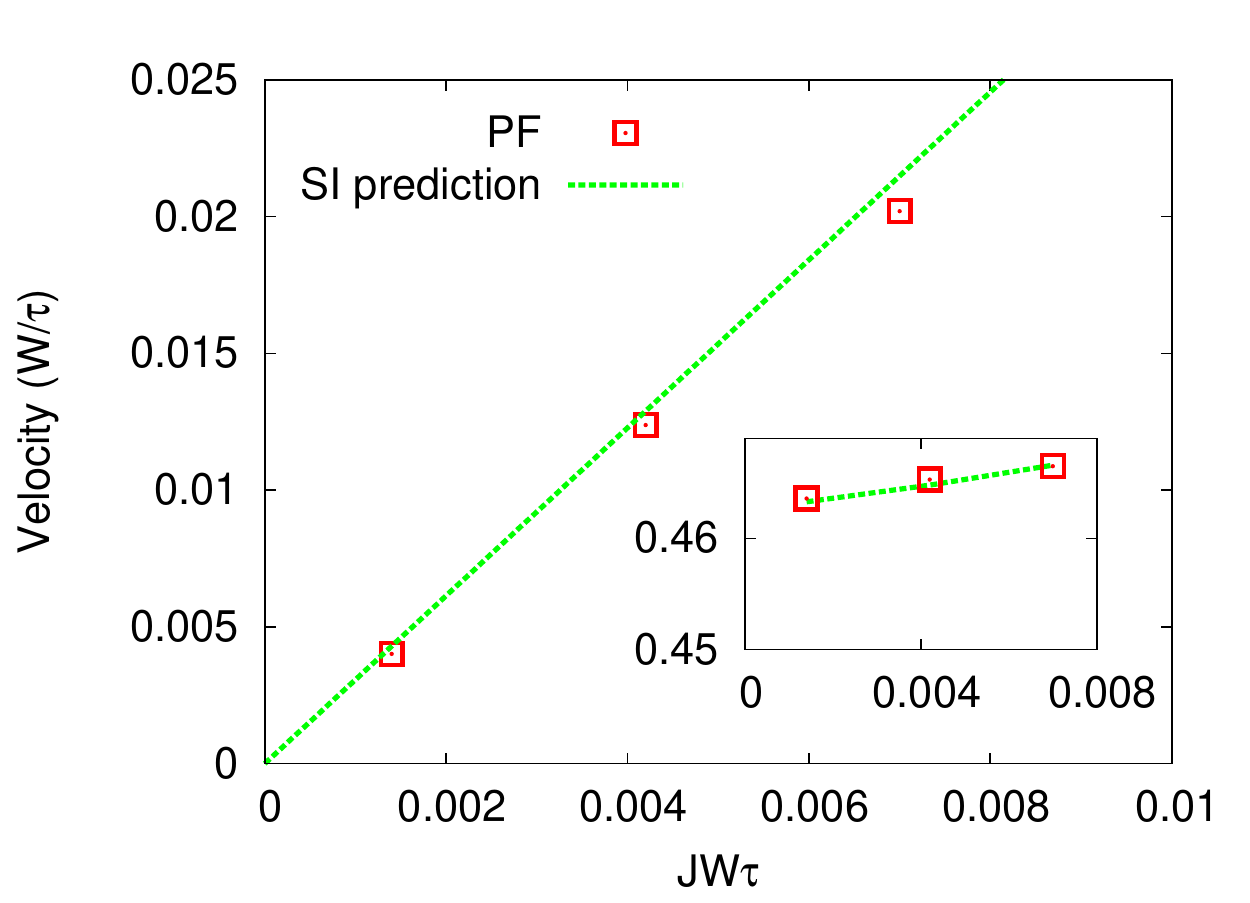}\\
  \caption{Comparison of the PF steady-state growth rate with the sharp-interface prediction (Eq.~\ref{v_J_eqn}).
  The inset is a comparison of the PF steady-state catalyst concentration with the sharp-interface prediction (Eq.~\ref{c_J_eqn}).
  The catalyst size is $A_0/W^2=313$.}
  \label{v_flux} \center
\end{figure}

For the second incorporation-rate-dominated regime, a simple
relation between the steady-state NW growth rate $V$ and the
incorporation flux $J$ can be derived from the wire geometry shown
in Fig. \ref{wire_angles} and the flux balance condition (Eq.~
\ref{growth_si_dndt})
\begin{equation}
V=J\Omega_s\frac{\pi-\theta_2}{\sin\theta_2},\label{v_J_eqn}
\end{equation}
The PF results agrees well with the prediction in Eq.~\ref{v_J_eqn}
as shown in Fig. \ref{v_flux}.
Eq.~\ref{v_J_eqn} can also be used to calculate the steady-state
concentration in the catalyst droplet by equating the growth
velocity in Eq.~\ref{v_J_eqn} to the sharp-interface velocity in
Eq.~\ref{growth_app_eqn_mpf_motion_sharpinterface}. Assuming
$\Omega_l=\Omega_s$ and using the solid-liquid interface mobility in
Eq.~\ref{M_sl_eqn_sum} together with Eq.~\ref{si_R_relations}, the
catalyst concentration is related to the sharp-interface NW radius
$R_{si}$ by
\begin{equation}
\beta(c_l-c_0)=J\frac{\gamma_{sl}\tau\Omega_s^2}{W^2}\frac{\pi-\theta_2}{\sin\theta_2}+\Omega_s\frac{\gamma_{sl}\sin\theta_1}{R_{si}}.\label{c_J_eqn}
\end{equation}
Combining Eq.~\ref{c_J_eqn} with the radius-volume relation in
Eq.~\ref{r_a_linear} and the volume-concentration relation in
Eq.~\ref{v_change_analytic}, one can predict the catalyst
concentration as a function of the incorporation flux $J$ as shown
in the inset of Fig.~\ref{v_flux}.

\begin{figure}[!h]
  \includegraphics[width=80mm]{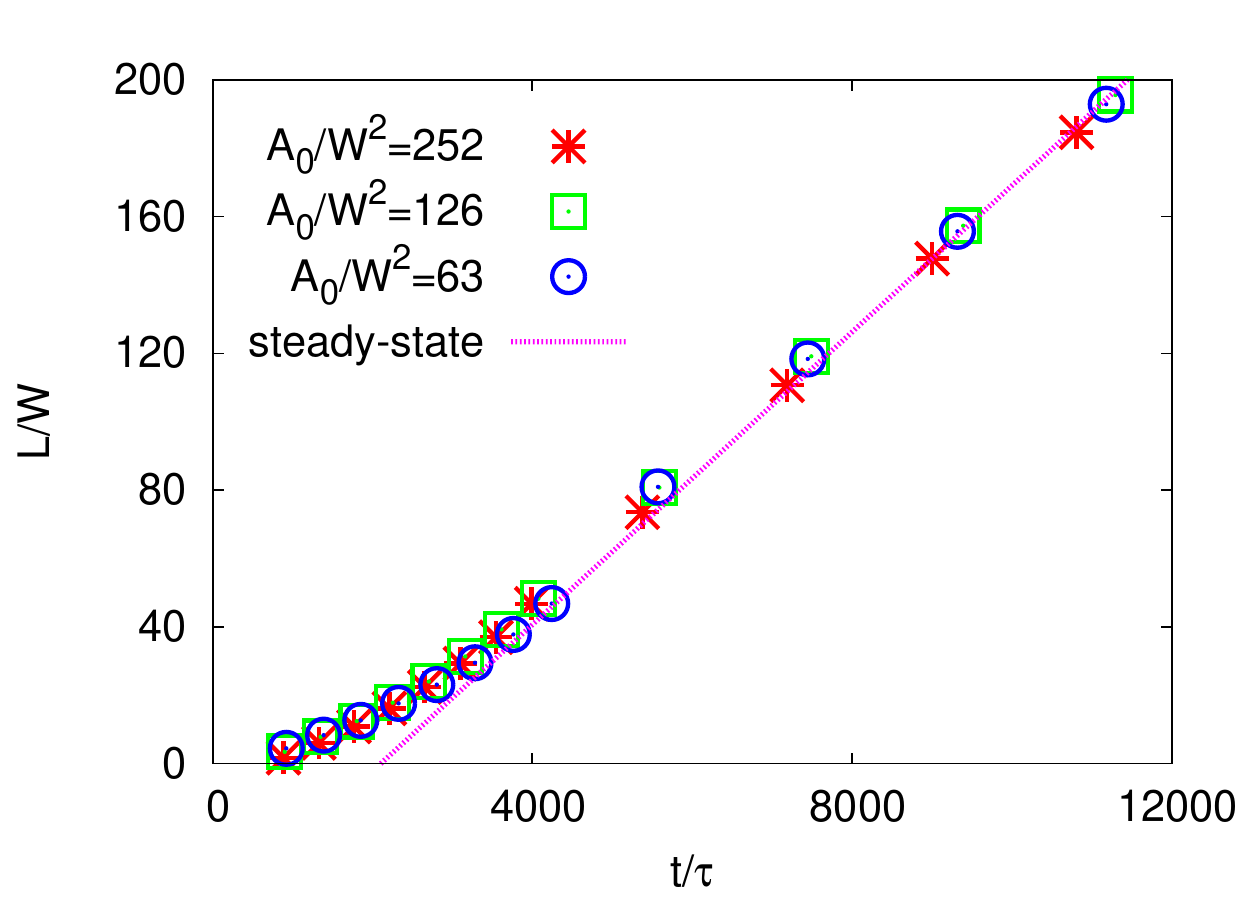}\\
  \caption{NW length $L$ as a function of time for different catalyst
  sizes in the incorporation-rate-dominated regime.
  Other parameters are the same as in Fig.~\ref{NW_grow_3step}.
  The steady-state line is based on predicted growth rate from Eq.~\ref{v_J_eqn}.}\label{l_flux}
\end{figure}

NWs of different size are also grown in this
incorporation-rate-dominated regime using the full $N_g$ dynamics
described in Eqs.~\ref{pf_dndt}, \ref{pf_mu} and \ref{pf_dvdt}.
Since the number of incorporated growth atoms into the droplet is
the product of the droplet surface area and a constant current
density $J$, the rate of incorporation of growth atoms into the solid scales as the
product of the NW growth rate $V$ and solid-liquid interface area
divided by the atomic volume of solid $\Omega_s$. Both the
droplet surface area and solid-interface area scale as $R^{d-1}$
(where $d$ is the spatial dimension), and $V\sim J\Omega_s$ becomes
independent of the NW radius in this limit, as shown in our
numerical results in Fig.~\ref{l_flux}.

\subsection{Faceted nanowire growth}

In this part we present results for faceted NW growth based on the
anisotropic PF model introduced in section \ref{secanis}. For
simplicity, we consider a solid-liquid anisotropy with $\gamma$-plot
of the form
\begin{equation}
\gamma_{sl}(\theta)=\gamma_{sl}^0 \frac{1+\delta_a|\sin2\theta|+\delta_b|\cos2\theta|}{1+\min(\delta_a,\delta_b)},\label{normalized_gamma}
\end{equation}
which has cusps at orientations $\theta=0,\pm \pi/2, \pi$ and
$\theta=\pm \pi/4, \pm 3\pi/4$ corresponding to $(10)$ and $(11)$
facets, respectively. Similar $\gamma$-plot has been measured
experimentally for Si~\cite{Si_anisotropy_exp} and computed for the
Si-Au system by atomistic simulations \cite{moneesh_md}. For the
liquid-vapor and the solid-vapor interfaces, we use isotropic form
$\gamma_{lv}=\gamma_{lv}^0$ and $\gamma_{sv}=\gamma_{sv}^0$.

To make $\gamma_{sl}(\theta)$ differentiable, we round the cusps by
replacing the absolute value function $|x|$ by a smooth function
$\sqrt{\epsilon^2+x^2}$, which transforms Eq.~\ref{normalized_gamma}
into a regularized form
\begin{equation}
\gamma_{sl}(\theta)=\gamma_{sl}^0\frac{1+\delta_a\sqrt{\sin^2
2\theta+\epsilon^2}+\delta_b\sqrt{\cos^2
2\theta+\epsilon^2}}{1+\min(\delta_a,\delta_b)}.\label{normalized_gamma_regular}
\end{equation}
This form is implemented in the PF model using the procedure
outlined in section \ref{secanis} and further detailed in appendix
\ref{growth_appendix_facet}. Simulations are carried out with
$\gamma_{sv}^0=1.2$ Jm$^{-2}$, $\gamma_{lv}^0=1.0$ Jm$^{-2}$,
$\gamma_{sl}^0=0.8$ Jm$^{-2}$, and $\epsilon=0.01$.

In principle, the anisotropy parameters $\delta_a$ and $\delta_b$
can be varied independently. For some regions of the ($\delta_a$
,$\delta_b$) parameter space, the solid-liquid interface stiffness
$\gamma_{sl}+d^2\gamma_{sl}/d\theta^2$ becomes negative over a range
of $\theta$ corresponding to thermodynamically unstable orientations
that are excluded from the equilibrium crystal shape. Those
so-called ``missing orientations'', defined by
$\gamma_{sl}+d^2\gamma_{sl}/d\theta^2<0$, are completely unrelated
to faceted orientations that, in contrast, correspond to large
positive extremal values of the stiffness in the regularized
$\gamma$-plot of the form of Eq. \ref{normalized_gamma_regular}. A
$\gamma$-plot, such as the one considered here and other more
general forms, can generally yield equilibrium shapes with missing
orientations and no facets, no missing orientations and facets, or a
mix of both facets and missing orientations. While various methods
have been developed to handle missing orientations in the PF model
\cite{sharp_corner1,sharp_corner2}, we restrict our attention here
to a region of the ($\delta_a$,$\delta_b$) parameter space that
yields solid-liquid equilibrium shapes without missing orientations.
For this purpose, we choose to constrain $\delta_a$ and $\delta_b$
by the relation $3\delta_a=\sqrt{1-9\delta_b^2}$, which is obtained
by requiring that the minimum value of the stiffness over all angles
equals zero, and hence that the stiffness is always equal to zero or
positive for $0\le \theta \le 2\pi$.  For
$\delta_a=\delta_b=\sqrt{2}/6$, the equilibrium shape is octagonal
and completely faceted with $(10)$ and $(11)$ facets of equal
lengths in the $\epsilon\rightarrow 0$ limit, while for $\delta_a\ne
\delta_b$, the equilibrium shape consists of facets of unequal
lengths and rough parts with finite positive stiffness. In the
latter case, $(10)$ facets have lower energy and are longer than
$(11)$ facets in the equilibrium shape for $\delta_a>\delta_b$ and
vice versa for $\delta_a<\delta_b$.

\begin{figure}[!h]
  \begin{center}
  \includegraphics[width=80mm]{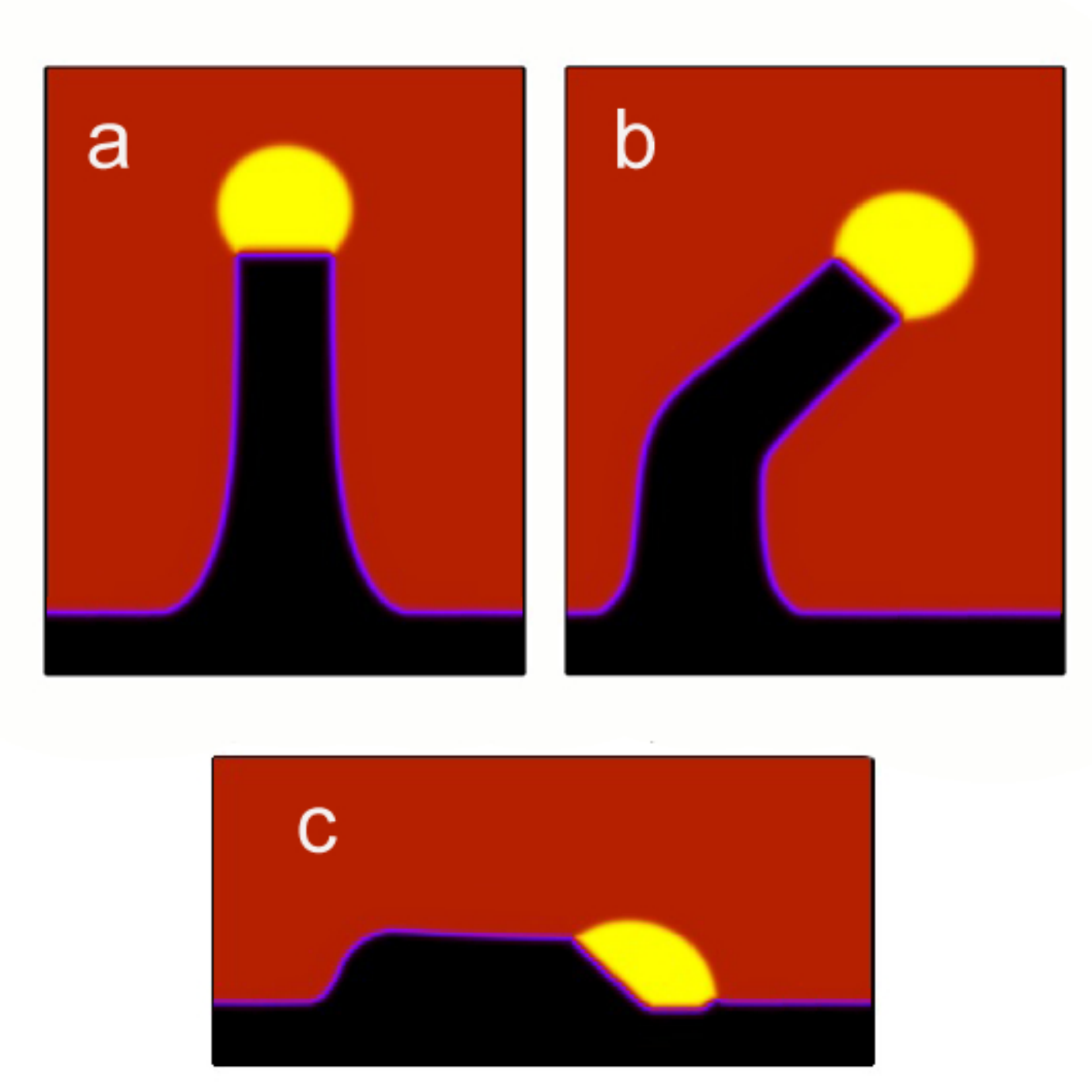}\\
  \caption{NW morphologies.
  Three phases are colored as red (vapor), yellow (liquid) and  black
  (solid). Catalyst size is $A_0/W^2=1254$. In (a),
  $\delta_a=0.3179,\delta_b=0.1$, $JW\tau=0.0035$.
  In (b), $\delta_a=0.1,\delta_b=0.3179$, $JW\tau=0.0035$ In (c), $\delta_a=0.1,\delta_b=0.3179$, $JW\tau=0.0007$.
  }\label{facet_morph}
  \end{center}
\end{figure}

Examples of NW growth from a $(10)$ substrate are shown in Fig.
\ref{facet_morph}. When $(10)$ facets are energetically favored
($\delta_a>\delta_b$), simulations reproduce the standard mode of
tapered growth normal to the substrate (Fig.~\ref{facet_morph}a). In
contrast, when $(11)$ facets are energetically favored
($\delta_a<\delta_b$), growth normal to the substrate becomes
unstable. For large enough flux of growth atoms (measured in our 2D
simulations by the dimensionless product $JW\tau$), the NW first
emerges normal to the substrate but then kinks towards another
direction after a finite growth distance, which corresponds to
$(11)$ in the example of Fig. \ref{facet_morph}b. In contrast, for
small flux, the NW is not able to emerge from the substrate before
kinking and instead crawls along the substrate as seen in Fig.
\ref{facet_morph}c.

\begin{figure}[!h]
  \begin{center}
  \includegraphics[width=80mm]{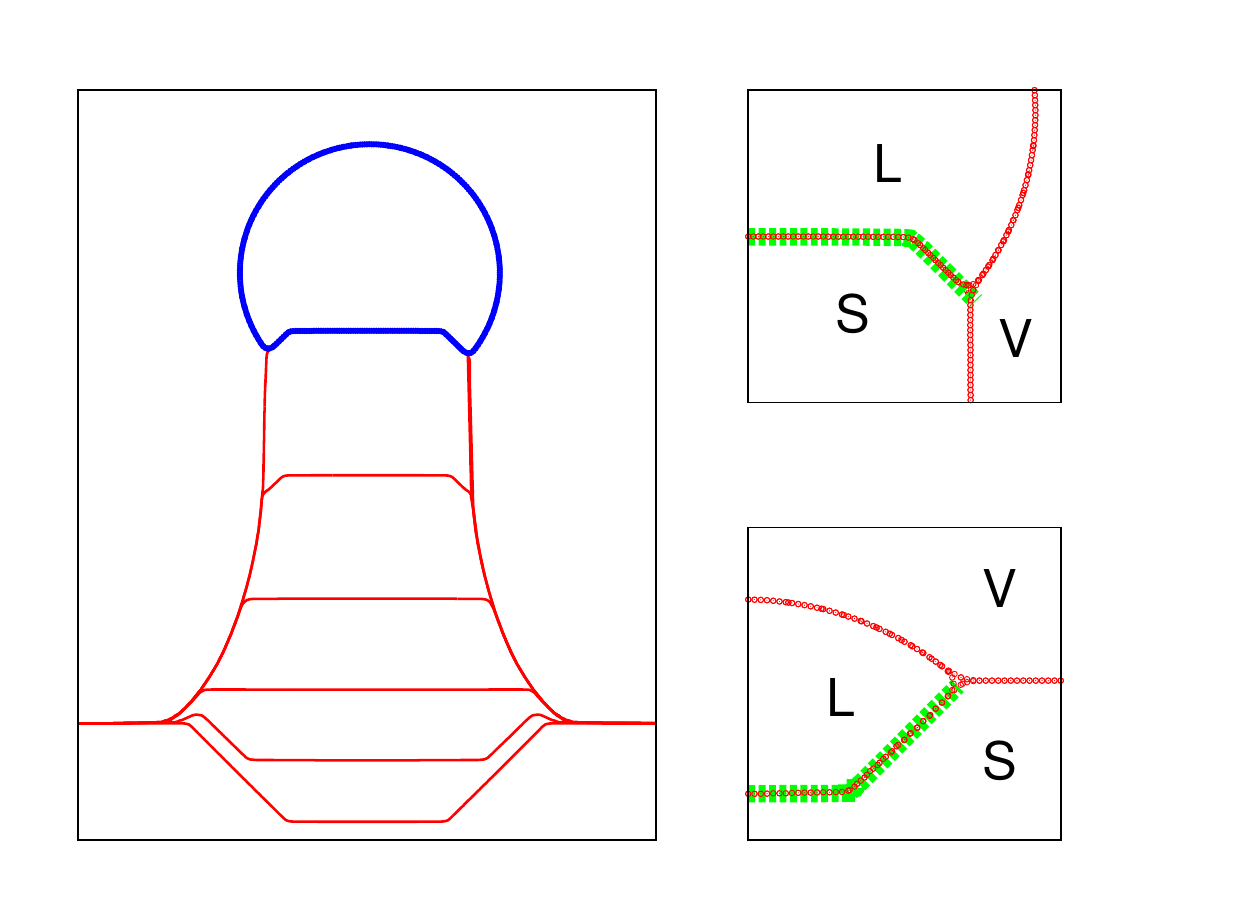}\\
  \caption{Phase-field simulation of NW growth from a $(10)$ substrate for $\delta_a=\delta_b=\sqrt{2}/6$, $JW\tau=0.0028$, and droplet size $A_0/W^2=1254$.
  Outlines of solid are shown at different stages of morphological development in the left panel (as red line).
  The blue lines depict the liquid-vapor and solid-liquid interfaces at the latest time where the NW is growing in steady-state.
  The phase-field solid-liquid interface shapes are compared to the prediction of sharp-interface theory (green dashed lines)
  during steady-state NW growth (top right inset) and for an
  equilibrium droplet on the substrate (bottom left inset). The three phases
  are labeled using their corresponding capital letter and all PF interfaces are shown as red lines in the two right
  panels. }\label{facet_eq}
  \end{center}
\end{figure}

Those simulations illustrate that NW growth is controlled by a
subtle balance of interface energetics and growth kinetics. A more
exhaustive study of NW growth behavior as a function of the
interface anisotropy parameters space, including solid-vapor
anisotropy that has been neglected here for simplicity, will be
presented elsewhere. In the rest of this section, we focus on
comparing the facetted NW tip shape obtained in phase-field
simulations to the one predicted by sharp-interface theory. For this
purpose, we focus on the case $\delta_a=\delta_b$ that yields a
completely faceted octagonal solid-liquid equilibrium crystal shape
with $(10)$ and $(11)$ facets of equal energies and equal lengths.
However, during NW growth from a $(10)$ substrate, the side facets
(i.e. $(\bar 11)$ and $(11)$ facets) are truncated to a shorter
length than the main $(10)$ facet, as illustrated by the phase-field
simulation in Fig. \ref{facet_eq}. This raises the question of how
to predict the length of truncated facets in the NW growth geometry.
Before addressing this question, we note that in the simulation of
Fig. \ref{facet_eq}, growth normal to the substrate is unstable with
isotropic solid-vapor interface. Normal growth was therefore
enforced by imposing zero flux boundary conditions on all phase
fields about a vertical axis that splits the NW into two equal
mirror symmetric parts. As will be described elsewhere, normal
growth can also obtained in a more physical way without imposing
mirror symmetry by making the solid-vapor interface faceted, with
facets modeled with a similar form of $\gamma$-plot as Eq.
\ref{normalized_gamma_regular}. However, the simulation of Fig.
\ref{facet_eq} suffices for the purpose of comparing the
steady-state NW growth shape to the prediction of sharp-interface
theory.

\begin{figure}[!h]
  \begin{center}
  \includegraphics[width=40mm]{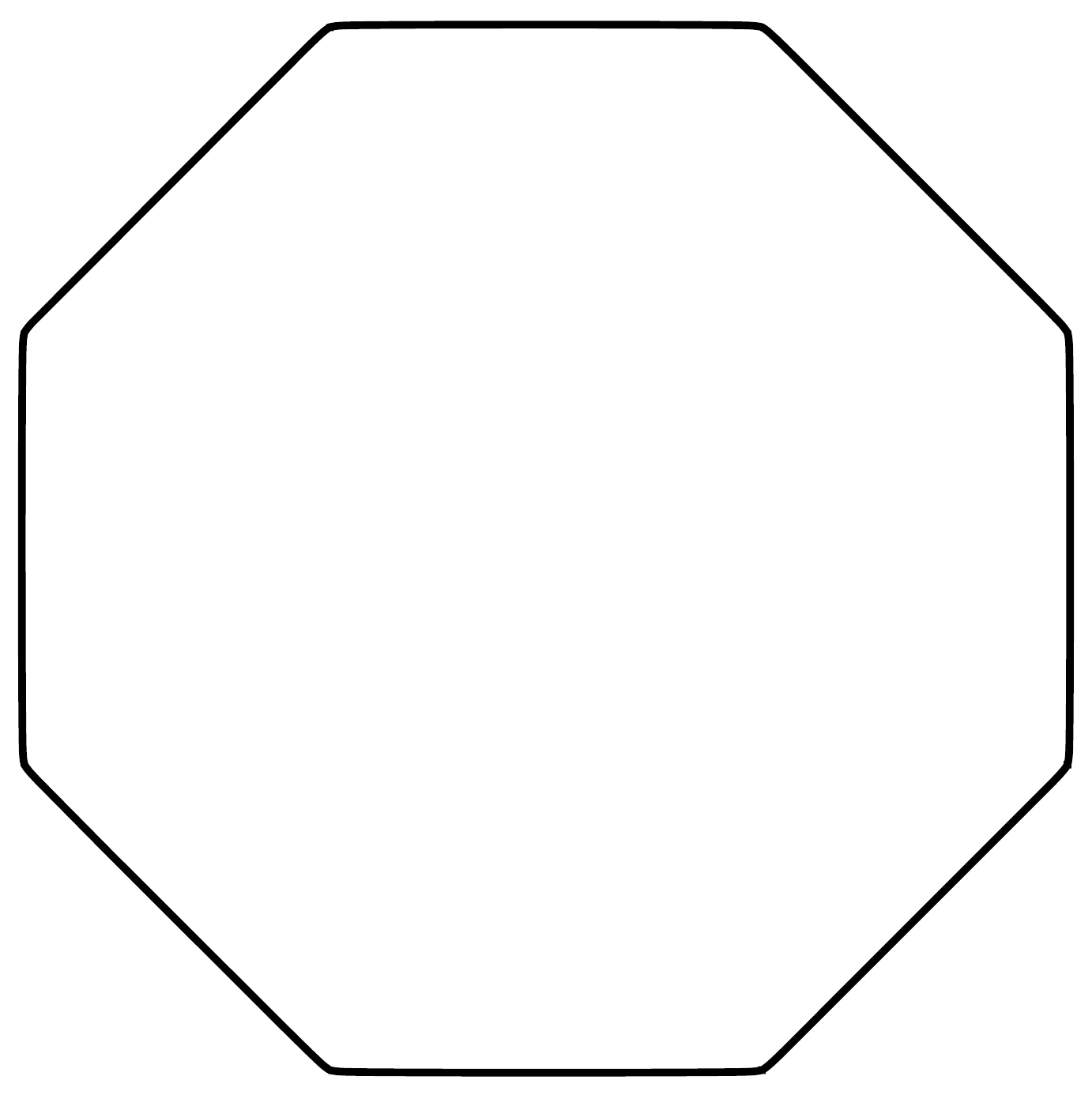}
    \includegraphics[width=40mm]{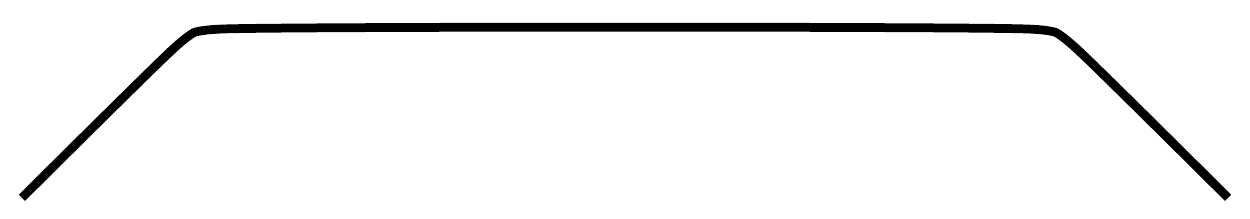}\\
  \caption{Computed solid-liquid interface shapes for a crystal seed surrounded by liquid (left) and for the steady-state tip shape of a growing NW in contact with a liquid droplet (right) where the side facets end at triple points. The parameters are $\delta_a=\delta_b=\sqrt{2}/6$, $\epsilon=0.01$,
  $\gamma_{sl}^0/\gamma_{lv}^0=0.8$, and $\gamma_{sv}^0/\gamma_{lv}^0=1.2$.}
  \label{eq_shapes}
  \end{center}
\end{figure}

\begin{figure}[!h]
  \begin{center}
  \includegraphics[width=70mm]{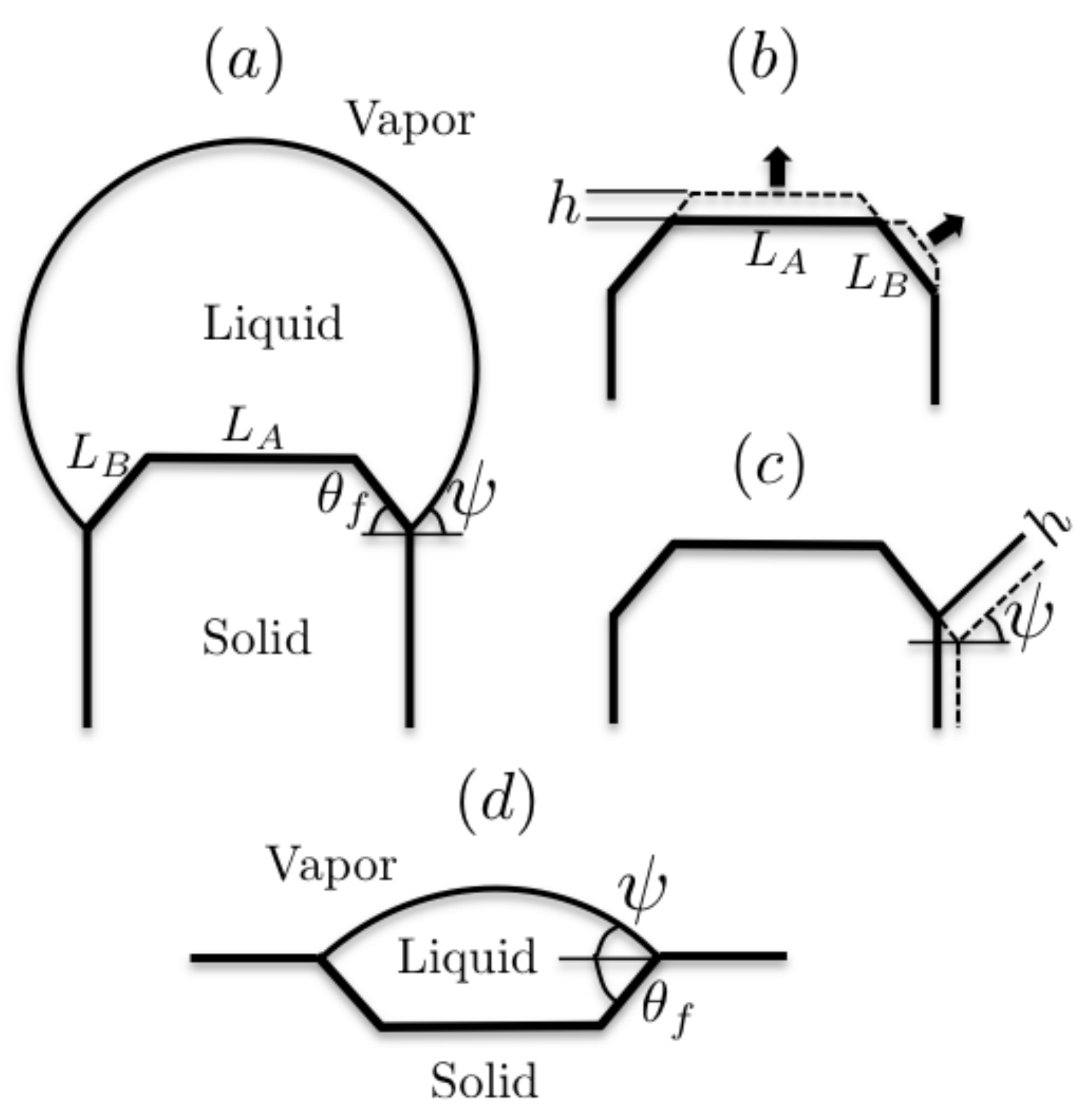}\\
  \caption{Schematic representation of (a) steady-state NW growth shape with a main facet of length $L_A$ and two truncated
  facets of length $L_B$, (b) virtual displacements of main and right truncated facets, (c) virtual displacement of right triple point, and (d)
  equilibrium droplet on substrate.}
  \label{virtual_fig}
  \end{center}
\end{figure}

The first approach is to apply the geometrical Wulff construction of
the equilibrium crystal shape. The latter can be expressed in an
equivalent parametric representation where the cartesian coordinates
of the interface are functions of $\theta$ given by \cite{facet_shape_voorhees}
\begin{eqnarray}
x(\theta)&=&\tilde\gamma_{sl}(\theta)\sin\theta + \tilde\gamma_{sl}'(\theta) \cos\theta \label{xeq}\\
y(\theta)&=&\tilde\gamma_{sl}(\theta)\cos\theta - \tilde\gamma_{sl}'(\theta) \sin\theta, \label{yeq}
\end{eqnarray}
where we have defined the dimensionless solid-liquid interface
energy $\tilde
\gamma_{sl}(\theta)=\gamma_{sl}(\theta)/\gamma_{lv}^0$. Here $x$ and
$y$ are taken to be dimensionless since the entire NW shape scales
proportionally to the NW diameter itself $\sim\sqrt{A}$. This
parametric representation is obtained as a solution of the
equilibrium Gibbs-Thomson condition
\begin{equation}
\left[\tilde\gamma_{sl}(\theta)+\tilde\gamma_{sl}''(\theta)\right]\kappa(\theta)=C,\label{gibbs_aniso}
\end{equation}
where $\kappa(\theta)$ is the interface curvature and $C$ is
constant. Eqs. \ref{xeq} and \ref{yeq} over the interval $0\le
\theta \le 2\pi$ define the equilibrium shape that is an octagon for
a crystal seed surrounded by liquid as shown in the left panel of
Fig. \ref{eq_shapes}. To compute the shape in the NW geometry, we
apply the anisotropic Young-Herring condition (\ref{Young-Herring})
at triple points. Projected onto the $x$ and $y$ axes, this
condition yields two independent equations
\begin{eqnarray}
\tilde\gamma_{sl}(\theta_f)\cos\theta_f =  \cos\psi + \tilde\gamma_{sl}'(\theta_f) \sin\theta_f \label{xproj},\\
 \tilde\gamma_{sl}(\theta_f)\sin\theta_f +  \tilde\gamma_{sl}'(\theta_f)\cos\theta_f+\sin \psi=\tilde\gamma_{sv}^0, \label{yproj}
\end{eqnarray}
respectively, where $\tilde\gamma_{sv}^0\equiv
\gamma_{sv}^0/\gamma_{lv}^0$. In addition, $\theta_f$ and $\psi$ are
the angles of the solid-liquid and solid-vapor interfaces measured
with respect to the horizontal axis as defined in Fig.
\ref{virtual_fig}a. The numerical solution of Eqs. \ref{xproj} and
\ref{yproj} with $\tilde \gamma_{sl}\equiv
\gamma_{sl}/\gamma_{lv}^0$ defined by Eq.
\ref{normalized_gamma_regular} uniquely determines $\theta_f$ and
$\psi$. The solid-liquid interface during NW growth is then the
parametric shape defined by Eqs. \ref{xproj} and \ref{yproj} for
$-\theta_f\le \theta \le \theta_f$. This shape is shown in the right
panel of Fig. \ref{eq_shapes}. It is also superimposed as a green
dashed line in the top-right panel of Fig. \ref{facet_eq} and seen
to agree well with the steady-state shape with truncated facets
predicted by phase-field simulations. Importantly, $\theta_f$ and
$\psi$ converge to unique values in the sharp cusp
$\epsilon\rightarrow 0$ limit. For $\epsilon=0.01$, those values are
almost reached. In particular, $\theta_f\approx 0.783$  is almost
$\pi/4$ expected of $(11)$ facets and $\psi\approx 0.832$ is also
very close to its $\epsilon\rightarrow 0$ limit derived below. A
similar calculation is straightforward to carry out for an
equilibrium droplet on a $(10)$ substrate. The predicted shape is
also in good quantitative agreement with the phase-field droplet
equilibrium shape in the bottom-right panel of Fig. \ref{facet_eq}
and qualitative agreement with experimental observations showing
dissolution of the substrate below the droplet \cite{Farralis}.

We now derive analytically the steady-state NW growth shape in the
sharp cusp limit ($\epsilon=0$). In this limit, the truncated $(\bar
11)$ and $(11)$ facets have fixed orientations such that $\theta_f=
\pi/4$ is no longer an unknown. The unknowns are $\psi$ and the
ratio $L_B/L_A$ of the truncated and main facet lengths. To
determine those unknowns, we consider small virtual displacements of
individual facets of fixed orientations and one of the triple points
that leave the total free-energy unchanged, as depicted in Fig.
\ref{virtual_fig}. The total free-energy change resulting from a
virtual displacement of the main $(10)$ facet a distance $h$ along
the direction normal to the facet (Fig. \ref{virtual_fig}b) is the
sum of bulk and interface contributions given by
\begin{equation}
\Delta f L_A h +\frac{2\gamma_Bh}{\sin\theta_f}-\frac{2\gamma_Ah}{\tan\theta_f}=0,\label{Afacet}
\end{equation}
where $\Delta f<0$ is the difference of free-energy density between
solid and liquid, $\gamma_A\equiv \gamma_{sl}(0)$ and
$\gamma_B\equiv \gamma_{sl}(\pi/4)$ are the facet free-energies, and
the second and third terms corresponds to the change of interface
free-energy resulting from the lengthening of the side $(\bar 11)$
and $(11)$ facets and the shortening of the main $(10)$ facet,
respectively. Similarly, the total free-energy change resulting from
the virtual displacement of the $(11)$ facet normal to itself
(Fig.\ref{virtual_fig}b) is given by
\begin{eqnarray}
& &\Delta f L_B h +\frac{\gamma_Ah}{\sin\theta_f}-\frac{\gamma_Bh}{\tan\theta_f}\nonumber \\
& &+\frac{(\gamma_{sv}^0-\gamma_{lv}^0\sin\psi)h}{\sin\left(\frac{\pi}{2}-\theta_f\right)}-\frac{\gamma_Bh}{\tan\left(\frac{\pi}{2}-\theta_f\right)}=0,\label{Bfacet}
\end{eqnarray}
and contains contributions from the changes of length of the $(10)$
and $(11)$ facets as well as the solid-vapor and liquid-vapor
interfaces. The virtual displacement of the right triple point
(Fig.\ref{virtual_fig}c) yields in turn
\begin{equation}
-\gamma_{sv}^0h\sin\theta_f+\gamma_Bh-h\cos(\theta_f+\psi)\gamma_{lv}^0=0.\label{triplepoint}
\end{equation}
Eliminating $\Delta f$ between Eqs. \ref{Afacet} and (\ref{Bfacet})
yields the prediction of the ratio of facet length
\begin{equation}
\frac{L_B}{L_A}=\frac{\gamma_A\cos\theta_f-\gamma_B+(\gamma_{sv}^0-\gamma_{lv}^0\sin\psi)\sin\theta_f}{2\cos\theta_f(\gamma_B-\gamma_A\cos\theta_f)}\label{lengthratio}
\end{equation}
with
\begin{equation}
\psi=\cos^{-1}\left(\frac{\gamma_B-\gamma_{sv}^0\sin\theta_f}{\gamma_{sv}^0}\right)-\theta_f,\label{psi}
\end{equation}
obtained from Eq. \ref{triplepoint}. For the parameters of the
simulations $\gamma_A=\gamma_B=\gamma_{sl}^0$,
$\gamma_{sl}^0/\gamma_{lv}^0=0.8$,
$\gamma_{sv}^0/\gamma_{lv}^0=1.2$, and $\theta_f=\pi/4$, Eq.
\ref{psi} predicts $\psi\approx 0.826$, which is close to the value
$\psi\approx 0.832$ predicted by the rounded-cusp approximation with
$\epsilon=0.01$. In addition, Eq. \ref{lengthratio} predicts
$L_B/L_A\approx 0.285$ that agrees well quantitatively with both
phase-field simulations and sharp interface theory with the rounded
cusp approximation.

Finally, the method of virtual displacement can also be used to
derive analogous analytical expressions for the ratio of facet
lengths and $\psi$ for an equilibrium droplet on a substrate, with
$\psi$ defined in Fig. \ref{virtual_fig}d.  The calculation is
straightforward and we only give here the final results
\begin{equation}
\frac{L_B}{L_A}=\frac{\gamma_A+\gamma_{lv}^0\cos\psi-\gamma_{sv}^0}{2(\gamma_B-\gamma_A\cos\theta_f)}\label{lengthratio_eq}
\end{equation}
with
\begin{equation}
\psi=\cos^{-1}\left(\frac{\gamma_{sv}^0\cos\theta_f-\gamma_B}{\gamma_{lv}^0}\right)-\theta_f,\label{psi_eq}
\end{equation}

\section{Conclusions and outlook}

%We developed a NW growth model based on the multi-phase-field
%method. It captures the evolution of NW from a catalyst seed to a
%fully developed wire. Comparing with traditional sharp-interface
%approach, our model is easy to implement and can be generalize to
%three dimensions with a reasonable computational cost. With
%information from atomistic calculations, this model has the
%potential to take the computational understanding of NW growth to a
%much longer time scale.
%
%In the development of this model, we also extended the phase-field
%theory in two ways. First, a novel sharp-interface asymptotics was
%derived for this model which reproduces the previously proposed
%sharp-interface NW growth model. Within this asymptotics, the idea
%of time scale separation at different interfaces played a crucial
%role in the recovering of correct sharp-interface limit. Second, an
%alternative approach to include surface energy anisotropy in
%multi-phase-field model based on the potential term in the free
%energy functional was developed and quantitatively validated.
%Comparing with the commonly used gradient-term based approach, this
%method is more flexible since its effect is only localized at the
%specified binary interface region and can be used to tune the
%anisotropy of different interfaces independently.

In summary, we have developed a multi-phase-field model to describe quantitatively
NW growth by the vapor-liquid-solid (VLS) process. This model uses a free-energy landscape similar to the one used previously to model eutectic solidification \cite{Plapp} and introduces several new features to adapt this model to the VLS system:
\begin{enumerate}
\item
The change of catalyst droplet volume, which is associated with the change of concentration of growth atoms inside the droplet, is described using a Lagrange multiplier, in addition to the Lagrange multiplier commonly used to constrain the sum of the phase fields to unity.
\item
The physically relevant limit of rapid equilibration of the liquid catalyst
to a droplet shape of constant mean curvature on the characteristic time scale of NW growth is achieved by choosing the liquid-vapor interface mobility much larger than the solid-liquid interface mobility. In this limit, the Lagrange multiplier used to constrain the volume of the catalyst reduces to the Laplace pressure inside the droplet, thereby providing a thermodynamically consistent description of the VLS system without the computational burden of treating the catalyst as a real fluid.
\item
The driving force for growth is incorporated by adding a non-variational term localized at the solid-liquid interface to the evolution equation for the solid and liquid phase fields, which is equivalent to lowering the free-energy of the solid with respect to the liquid on this binary interface.
\item
Anisotropy of the excess interfacial free-energy is introduced by making the free-energy barrier height between two phases dependent on interface orientation.
\end{enumerate}
We have presented a detailed asymptotic analysis of the model in the limit where the interface thickness is much smaller than the NW radius and shown that phase-field equations reduce in this limit to a previously proposed sharp-interface model of
NW growth by Schwarz and Tersoff \cite{Tersoff_PRL2009}. The
simulations reproduce the complex evolution of the interfaces from a droplet on a substrate to steady-state NW growth normal to the substrate with tapering of the side walls. Furthermore, the model can describe different experimentally observed growth regimes including the regime where the growth rate is limited by the solid-liquid interface kinetics, in which case the growth rate depends on the NW radius, and the opposite regime where the growth rate is limited by the incorporation rate of growth atoms at the catalyst surface, in which case the growth rate is independent of radius.

With the incorporation of an anisotropic solid-liquid $\gamma$-plot that contains faceted interfaces, the model can also reproduce the characteristic solid-liquid interface NW tip shape consisting of a main facet
intersected by two truncated side facets ending at triple points, as well as more complex growth behaviors including NW kinking and crawling. Finally, we have developed a sharp-interface theory to predict the length of the main facet and truncated facet and shown that the predictions are in good agreement with phase-field simulations.

While the simulations presented in this paper were restricted to two dimensions, the present PF model can be readily implemented in three dimensions to carry out a quantitative comparison with experimentally observed NW growth shapes. Three-dimensional simulation results in the Si-Au VLS system will be presented elsewhere. Another interesting prospect is to extend the proposed theoretical description of faceted NW growth shape to three dimensions.

\section{acknowledgement}
This research was supported by the National Science Foundation grant 1106214 from the DMR CMMT Program.

\appendix
\section{Derivation of the Lagrange multiplier controlling the catalyst volume}
\label{growth_appendix_lag} In this PF NW growth model, the catalyst
size constraint is incorporated by adding a Lagrange multiplier
$\lambda_A$ to the original free energy functional $F$
\begin{equation}
\tilde{F}=F-\lambda_Ah\left[\int g_l(\vec{\phi})dv-A(t)\right],
\end{equation}
where $A$ is the droplet volume at time t and $g_l$ is a
function given in the main text that varies smoothly between 1 in the liquid and 0 in other phases such that
$\int g_l(\vec{\phi})dv$ measures the total droplet volume. Minimization of $\tilde{F}$ with respect to
$\phi$ gives a configuration where the total catalyst volume
(measured by the integral of $g_l$) is constrained to be $A(t)$. The
evolution equation derived from $\tilde{F}$ is
\begin{equation}
\tau\frac{\partial \phi_{i}}{\partial
t}=-\bar{K}(\vec{\phi})h^{-1}\left(\frac{\hat{\delta}
F}{\hat{\delta} \phi_i}-\lambda_Ah\frac{\partial g_l}{\partial
\phi_i}\right),\label{growth_app_b2}
\end{equation}
where the modified functional derivative $\frac{\hat{\delta}
F}{\hat{\delta} \phi_i}$ are defined by Eqs.~\ref{del_F_hat1} to
\ref{del_F_hat3}. The other constraint $\sum_{i=1}^3\phi_i=1$ can be
included by adding a summation term to the equation of
motion~\cite{Plapp}
\begin{equation}
\tau\frac{\partial \phi_{i}}{\partial
t}=-K(\vec{\phi})h^{-1}\left(\frac{\hat{\delta}
\tilde{F}}{\hat{\delta}
\phi_i}-\frac{1}{3}\sum_{j=1}^3\frac{\hat{\delta}
\tilde{F}}{\hat{\delta} \phi_j}\right). \label{growth_app_b3}
\end{equation}

A given droplet volume evolution
\begin{equation}
\frac{dA}{dt}=\frac{d}{dt}\int
g_l(\vec{\phi})dv=\dot{A},\label{growth_app_b3.5}
\end{equation}
can be written as
\begin{equation}
\int \frac{\partial
g_l(\vec{\phi})}{\partial t}dv=\int\sum_{i=1}^2\left(\frac{\partial
\tilde{g_l}}{\partial \phi_i}\frac{\partial \phi_i}{\partial
t}\right)dv, \label{growth_app_b4}
\end{equation}
by moving the time derivative inside the integral. Since the phase
fraction condition $\sum_{i=1}^3\phi_i=1$ is enforced in the
dynamics, $g_l$ is replaced by $\tilde{g_l}$ that depends only on two phase fields.
Using $\partial \phi_i/\partial t$ in Eq.~\ref{growth_app_b3}, one
can rewrite Eq.~\ref{growth_app_b4} as
\begin{eqnarray}
&&\nonumber\dot{A}\tau=-\int Kh^{-1}\Bigg\{\sum_{i=1}^2
\frac{\hat{\delta} F}{\hat{\delta} \phi_i}\frac{\partial
\tilde{g_l}}{\partial \phi_i}-\frac{1}{3}\sum_{i=1}^2 \frac{\partial
\tilde{g_l}}{\partial
\phi_i}\sum_{j=1}^3\frac{\hat{\delta} F}{\hat{\delta} \phi_j}\\
&&-\lambda_Ah\left[ \sum_{i=1}^2 \left(\frac{\partial
\tilde{g_l}}{\partial\phi_i}\right)^2-\frac{1}{3}\left(\sum_{i=1}^2
\frac{\partial \tilde{g_l}}{\partial
\phi_i}\right)^2\right]\Bigg\}dv,\label{growth_app_b5}
\end{eqnarray}
where $g_l$ in $\tilde{F}$ is also replaced by $\tilde{g_l}$.
Solving for $\lambda_A$ using Eq.~\ref{growth_app_b5} gives
\begin{equation}
\lambda_A=\frac{I_1- I_2+\dot{A}\tau}{I_3-I_4},\label{growth_app_b6}
\end{equation}
with
\begin{equation}
I_1=\int Kh^{-1}\sum_{i=1}^2 \frac{\hat{\delta} F}{\hat{\delta}
\phi_i}\frac{\partial \tilde{g_l}}{\partial \phi_i}dv,
\end{equation}
\begin{equation}
I_2=\frac{1}{3}\int Kh^{-1}\sum_{i=1}^2 \frac{\partial
\tilde{g_l}}{\partial \phi_i}\sum_{j=1}^3\frac{\hat{\delta}
F}{\hat{\delta} \phi_j}dv,
\end{equation}
\begin{equation}
I_3=\int K\sum_{i=1}^2\left(\frac{\partial
\tilde{g_l}}{\partial\phi_i}\right)^2dv,
\end{equation}
\begin{equation}
I_4=\frac{1}{3}\int K\left(\sum_{i=1}^2 \frac{\partial
\tilde{g_l}}{\partial \phi_i}\right)^2dv.
\end{equation}

\section{Sharp-interface limit for an isolated droplet with volume constraint}
\label{growth_si_appendix}

The sharp-interface limit of a single order-parameter PF model with
a volume-controlling Lagrange multiplier is worked out in this
section. This case corresponds physically to an isolated liquid droplet inside the vapor phase.
The main purpose of this appendix is to
understand the role of this Lagrange multiplier $\lambda_A$ in this simpler setting.
The same formulation is used to control the catalyst
size in our multiphase-field VLS NW growth model. The sharp-interface limit
of this model is examined in
section~\ref{growth_si} using results derived in this appendix.

The free-energy is given by
\begin{equation}
F_1=\int
h\left[\frac{W^2}{2}|\nabla\phi|^2+f(\phi)\right]dv,\label{growth
app_simplepf_F}
\end{equation}
with $f(\phi)$ being a double-well potential which has two minima at
$\phi=0$ and $\phi=1$ corresponding to the vapor and liquid phase, respectively.
The evolution equation for the phase field $\phi$ is
\begin{equation}
\tau\frac{\partial \phi}{\partial t}=-K_1(\phi)h^{-1}\left(\frac{\delta F_1}{\delta
\phi}-\lambda_A\frac{\partial g}{\partial
\phi}\right),
\label{growth_app_eqn_simplepf}
\end{equation}
where $g$ is a smooth tilt function (similar to $g_l$ in the
multiphase-field model) that varies smoothly between 0 and 1, $K_1(\phi)$ is the dimensionless mobility (similar to
$K(\vec{\phi})$ in the multiphase-field model), and
\begin{equation}
\lambda_A=\frac{\int K_1(\phi)h^{-1}\frac{\delta F_1}{\delta
\phi}\frac{\partial g}{\partial \phi}dv+\dot{A}\tau}{\int
K_1\left(\frac{\partial g}{\partial \phi}\right)^2dv},
\label{growth_app_eqn_simplepf_lag}
\end{equation}
is a volume-controlling Lagrange multiplier,
which is derived using the approach outlined in appendix
\ref{growth_appendix_lag}. Using Eq.~\ref{growth app_simplepf_F},
Eq.~\ref{growth_app_eqn_simplepf} becomes
\begin{equation}
\tau\frac{\partial \phi}{\partial
t}=K_1\left[W^2\nabla^2\phi-\frac{\partial f}{\partial
\phi}+\lambda_A\frac{\partial g}{\partial \phi}\right].
\label{growth_app_eqn_simplepf_2}
\end{equation}

To get the corresponding sharp-interface limit of
Eq.~\ref{growth_app_eqn_simplepf_2}, a sharp-interface analysis is
carried out in the following. Unlike the method used in the
sharp-interface expansion of PF solidification models where both
outer and inner expansions are performed \cite{Alain_quantitativePF_PRE,Alain_quantitativePF_PRL}, only an inner
expansion on the scale of the interface thickness is needed here since the interface dynamics is not controlled by a long range
diffusion field.

In order to characterize the motion of the interface, we define
the local curvilinear coordinate system ($r,s$) where $r(x,y,t)$ and $s(x,y,t)$ measure the position along
a direction $\hat{r}$ normal to the interface (where $\hat{r}$ points to the $\phi=0$ phase)
and along a direction $\hat{t}$ parallel to constant $\phi$ contours (along the interface) in a frame moving at the normal velocity of the interface.
In this coordinate system, the
$\nabla^2\phi$ term in Eq.~\ref{growth_app_eqn_simplepf_2}
reduces to
\begin{equation}
\nabla^2\phi=\frac{\partial^2\phi}{\partial
r^2}+\kappa\frac{\partial\phi}{\partial r}+(\nabla
s)^2\frac{\partial^2\phi}{\partial
s^2}+\nabla^2s\frac{\partial\phi}{\partial
s},\label{growth_app_x_transform}
\end{equation}
where $\nabla^2r=\kappa$ and $|\nabla r|=1$ are used in derivation.
Since $(r,s)$ are defined in a moving frame, the time derivative
$\partial\phi/\partial t$ in Eq.~\ref{growth_app_eqn_simplepf_2} is
replaced by
\begin{equation}
\frac{\partial\phi}{\partial t}\rightarrow
\frac{\partial\phi}{\partial t}+\frac{\partial r}{\partial
t}\frac{\partial \phi}{\partial r}+\frac{\partial s}{\partial
t}\frac{\partial \phi}{\partial s}.\label{growth_app_t_transform}
\end{equation}
To study the motion of $\phi$ in the sharp-interface limit, a
mesoscopic length $l_c$ is introduced such that the interface
thickness $W$ is small comparing with $l_c$ or
$p=W/l_c\rightarrow0$. Rescaling
Eq.~\ref{growth_app_eqn_simplepf_2} with length scale $l_c$ and time
scale $l_c^2/D$ (where $D$ has the dimension of interface mobility
$M$ times the surface energy $\gamma$), we obtain
\begin{equation}
\alpha p^2\frac{\partial \phi}{\partial
t}=p^2\nabla^2\phi-\frac{\partial f}{\partial
\phi}+\lambda_A\frac{\partial g}{\partial \phi},
\label{growth_app_eqn_simplepf_3}
\end{equation}
where $\alpha=\tau D/(W^2 K_1)$. Using
Eqs.~\ref{growth_app_x_transform} and \ref{growth_app_t_transform},
Eq.~\ref{growth_app_eqn_simplepf_3} becomes
\begin{eqnarray}
\nonumber\alpha p^2\left(\frac{\partial \phi}{\partial
t}-\frac{\partial\phi}{\partial r}v+\frac{\partial\phi}{\partial
s}\frac{\partial s}{\partial
t}\right)&=&p^2\left[\frac{\partial^2\phi}{\partial
r^2}+\kappa^0\frac{\partial\phi}{\partial r}\right.\\
\nonumber&&\left.+(\nabla s)^2\frac{\partial^2\phi}{\partial
s^2}+\nabla^2s\frac{\partial\phi}{\partial
s}\right]\\
&&-\frac{\partial f}{\partial \phi}+\lambda_A\frac{\partial
g}{\partial \phi}, \label{growth_app_eqn_simplepf_4}
\end{eqnarray}
where $v=-\partial r/\partial t$ and $\kappa^0=\kappa l_c$ is the
scaled curvature. Next, a stretched variable $z=r/p$ is introduced
such that it maps the interface region into $(-\infty,+\infty)$ in
$z$. Using this new variable $z$,
Eq.~\ref{growth_app_eqn_simplepf_4} becomes
\begin{eqnarray}
\nonumber\alpha p^2\left(\frac{\partial \phi}{\partial
t}-\frac{1}{p}\frac{\partial\phi}{\partial
z}v+\frac{\partial\phi}{\partial s}\frac{\partial s}{\partial
t}\right)&=&p^2\left[\frac{1}{p^2}\frac{\partial^2\phi}{\partial
z^2}+\kappa^0\frac{1}{p}\frac{\partial\phi}{\partial z}\right.\\
\nonumber&&\left.+(\nabla s)^2\frac{\partial^2\phi}{\partial
s^2}+\nabla^2s\frac{\partial\phi}{\partial
s}\right]\\
&&-\frac{\partial f}{\partial \phi}+\lambda_A\frac{\partial
g}{\partial \phi}. \label{growth_app_eqn_simplepf_5}
\end{eqnarray}
Keeping only $O(p)$ terms, Eq.~\ref{growth_app_eqn_simplepf_5} is
reduced to
\begin{equation}
-\alpha p \phi_z
v=\phi_{zz}+p\kappa^0\phi_z-q(\phi)+\lambda_A\frac{\partial
g}{\partial \phi}, \label{growth_app_eqn_simplepf_6}
\end{equation}
where $\phi_z$ and $\phi_{zz}$ are the first and the second
derivative of $\phi$ with respect to $z$, and $q(\phi)=\partial
f/\partial \phi$. All $s$ related terms are dropped since they are
all of $O(p^2)$. The phase field $\phi$ can also be expanded
in the small parameter $p$
\begin{equation}
\phi=\phi^0+p\phi^1+p^2\phi^2+\ldots.
\end{equation}
With this expansion, Eq.~\ref{growth_app_eqn_simplepf_6} is further
reduced to
\begin{eqnarray}
\nonumber-\alpha p \phi^0_z
v&=&\phi^0_{zz}+p\phi^1_{zz}+p\kappa^0\phi^0_z-q(
\phi^0)-q'(\phi^0)p\phi^1\\
&&+\lambda_Ag'(\phi^0), \label{growth_app_eqn_simplepf_7}
\end{eqnarray}
where $q'$ and $g'$ denote the derivatives of $q$ and $g$ with
respect to $\phi$, respectively. Since the interface
velocity $v$ only appears in the $O(p)$ term on the left-hand-side
of Eq.~\ref{growth_app_eqn_simplepf_7}, stationary interface
properties of the model are given by O(1) terms in
Eq.~\ref{growth_app_eqn_simplepf_7}. Since, furthermore, the Lagrange multiplier
$\lambda_A$ is used to control liquid volume in dynamics, it should
only appear at the same (or higher) order of the interface velocity,
i.e. $\lambda_A=p\lambda_A^1+O(p^2)$). To $O(1)$,
Eq.~\ref{growth_app_eqn_simplepf_7} becomes
\begin{equation}
\phi^0_{zz}-q(\phi^0)=0. \label{growth_app_eqn_simplepf_exp1}
\end{equation}
where $f=\phi^2(1-\phi)^2$ gives $q(\phi)=2\phi(1-\phi)(1-2\phi)$.
Solution of Eq.~\ref{growth_app_eqn_simplepf_exp1} gives the
stationary phase boundary profile
\begin{equation}
\phi^0(z)=\frac{1}{2}\left[1-\tanh\left(\frac{z}{\sqrt{2}}\right)\right].
\end{equation}
To $O(p)$, Eq.~\ref{growth_app_eqn_simplepf_7} is
\begin{equation}
-(\alpha
v+\kappa^0)\phi^0_z-\lambda_A^1g'(\phi^0)=\phi^1_{zz}-q'(\phi^0)\phi^1.
\label{growth_app_eqn_simplepf_exp2}
\end{equation}
Defining a linear operator
\begin{equation}
L\equiv\frac{\partial^2}{\partial z^2}-q'(\phi^0),
\end{equation}
Eq.~\ref{growth_app_eqn_simplepf_exp2} can be rewritten as
\begin{equation}
L\phi^1=-(\alpha
v+\kappa^0)\phi^0_z-\lambda_A^1g'(\phi^0).\label{si_eigen_phi1}
\end{equation}
Eq.~\ref{growth_app_eqn_simplepf_exp1} implies that $\phi^0_z$ is
a zero mode of the linear operator $L$ (eigenfunction with zero eigenvalue) since
\begin{equation}
L\phi^0_z=\phi^0_{zzz}-q'(\phi^0)\phi^0_z=0.
\end{equation}
Furthermore, since $L$ is self-adjoint, the right-hand-side of Eq.~\ref{si_eigen_phi1} must be orthogonal to the null space of $L$ for a nontrivial solution of Eq.~\ref{si_eigen_phi1} to exist,  which yields the standard solvability condition
\begin{equation}
\int^{+\infty}_{-\infty}\phi^0_z\left[-(\alpha
v+\kappa^0)\phi^0_z-\lambda_A^1g'(\phi^0)\right]dz=0,
\label{growth_app_eqn_simplepf_exp3}
\end{equation}
which can be further simplified to
\begin{equation}
v=\frac{-\kappa^0\gamma^0+\lambda_A^1}{Q},
\label{growth_app_eqn_simplepf_exp4}
\end{equation}
with $\gamma^0=\int^{+\infty}_{-\infty}(\phi^0_z)^2dz$ and
$Q=\int^{+\infty}_{-\infty}\alpha(\phi^0_z)^2dz$. By replacing $v$
and $\kappa^0$ with their unscaled dimensional form $v=V l_c/D$,
$\kappa^0=l_c\kappa$, and using the dimensional surface energy
$\gamma=Wh\gamma^0$, Eq.~\ref{growth_app_eqn_simplepf_exp4} becomes
\begin{equation}
V=M(-\kappa\gamma+\lambda_Ah), \label{growth_app_eqn_simplepf_exp5}
\end{equation}
with mobility
\begin{equation}
M=\frac{W}{\tau
h}\frac{1}{\int^{+\infty}_{-\infty}(\phi^0_z)^2K_1^{-1} dz}.
\end{equation}

Up to now, the sharp interface expression of $\lambda_A$ is still
unknown. Using the volume control condition
(Eq.~\ref{growth_app_b3.5}) with $g_l(\vec{\phi})$ replaced by
$g(\phi)$ and the coordinate transform in
Eq.~\ref{growth_app_t_transform}, we have
\begin{equation}
\int g'\left(\frac{\partial\phi}{\partial t}-V\frac{\partial
\phi}{\partial r}+\frac{\partial \phi}{\partial s}\frac{\partial
s}{\partial t}\right)dsdr=\dot{A},\label{app_lag_derive_eq1}
\end{equation}
where the volume integral has been replaced by
\begin{equation}
dv\rightarrow ds dr,
\end{equation}
with elements of arclength and radial coordinates $ds$ and $dr$, respectively.
It is important to note that the dimensional velocity $V$ is used here since time and
space are not rescaled. With the stretched coordinate transform
$z=r/p$, keeping only the leading order terms on the left-hand-side
gives
\begin{equation}
-\int g'\phi_zVdsdz=\dot{A}.\label{app_volume_change}
\end{equation}
Replacing $V$ with Eq.~\ref{growth_app_eqn_simplepf_exp5}, the
Lagrange multiplier is then
\begin{equation}
\lambda_Ah=\frac{\dot{A}}{MS}+\frac{\gamma\int\kappa ds}{S},
\label{growth_app_eqn_simplepf_lag_res}
\end{equation}
where $S=\int ds$ is the length of the interface. Using
Eq.~\ref{growth_app_eqn_simplepf_lag_res},
Eq.~\ref{growth_app_eqn_simplepf_exp5} becomes
\begin{equation}
V=M\left(-\kappa\gamma+\gamma\frac{\int\kappa
ds}{S}\right)+\frac{\dot{A}}{S},
\label{growth_app_eqn_simplepf_final_2}
\end{equation}
which is the droplet evolution Eq. \ref{modified_Cahn_Allen}.

\section{Incorporation of anisotropic solid-liquid interfacial free-energy and facets}
\label{growth_appendix_facet}

We consider a solid-liquid $\gamma$-plot of the form
energy function
\begin{equation}
\gamma_{sl}(\theta)=\gamma^0[1+\delta_a|\sin2\theta|+\delta_b|\cos2\theta|],\label{simple_gamma}
\end{equation}
where $\theta$ is the angle of the interface normal direction with respect to a reference crystal axis. According to the Wulff construction,
this $\gamma$-plot yields an equilibrium crystal shape with two sets of (10) and (11) facets at that are shown in the left of
Fig.~\ref{eq_shapes}.

To incorporate the interface free-energy anisotropy in Eq.~\ref{simple_gamma} into
the multiphase-field model, we treat $a_i$ in Eq.~\ref{gamma_general} as an orientation-dependent
parameter $a_i(\theta)$ where $\theta$ is the interface orientation
angle. Since there are two PF variables ($\phi_j$ and $\phi_k$)
involved at a binary interface in this model, the interface orientation can
be expressed using either
\begin{equation}
\sin\theta_j=-\partial_y\phi_j/|\nabla\phi_j|,
\end{equation}
or
\begin{equation}
\sin\theta_k=-\partial_y\phi_k/|\nabla\phi_k|.
\end{equation}
Since the barrier term $f_a^i$ is symmetric under the exchange of
$\phi_j$ and $\phi_k$, the same property should hold for
$a_i(\theta)$. A simple choice is then
\begin{equation}
a_i(\theta)=\left[\frac{1}{2}a_i(\theta_j)+\frac{1}{2}a_i(\theta_k)\right],\label{aniso_ai}
\end{equation}
which averages the contribution from both $\phi_j$ and $\phi_k$.
From here, the functional derivative is given by
\begin{eqnarray}
\frac{1}{h}\frac{\delta F}{\delta \phi_i}&=&\frac{\partial
f_d^i}{\partial
\phi_i}-W^2\nabla^2\phi_i+\sum_{l=1}^3\Bigg[a_l\frac{\partial
f_a^l}{\partial \phi_i}+b_l\frac{\partial f_b}{\partial
\phi_i}\\
\nonumber&&+\frac{\partial}{\partial
x}\left(\frac{\phi_{i,y}}{|\nabla\phi_i|^2}
f_a^la_{l,i}\right)-\frac{\partial}{\partial
y}\left(\frac{\phi_{i,x}}{|\nabla\phi_i|^2}f_a^la_{l,i}\right)\Bigg],
\end{eqnarray}
where we have defined
\begin{equation}
a_{l,i}=\frac{1}{2}\frac{\partial a_l(\theta_i)}{\partial
\theta_i}, \phi_{i,x}=\frac{\partial \phi_i}{\partial x}.
\end{equation}
In the numerical implementation, the orientation dependent terms are
only calculated in the interface region that is defined by $|\nabla\phi_i|<\epsilon_i$ where $\epsilon_i$ is a small cutoff.

\begin{figure}
  \begin{center}
  \includegraphics[width=80mm]{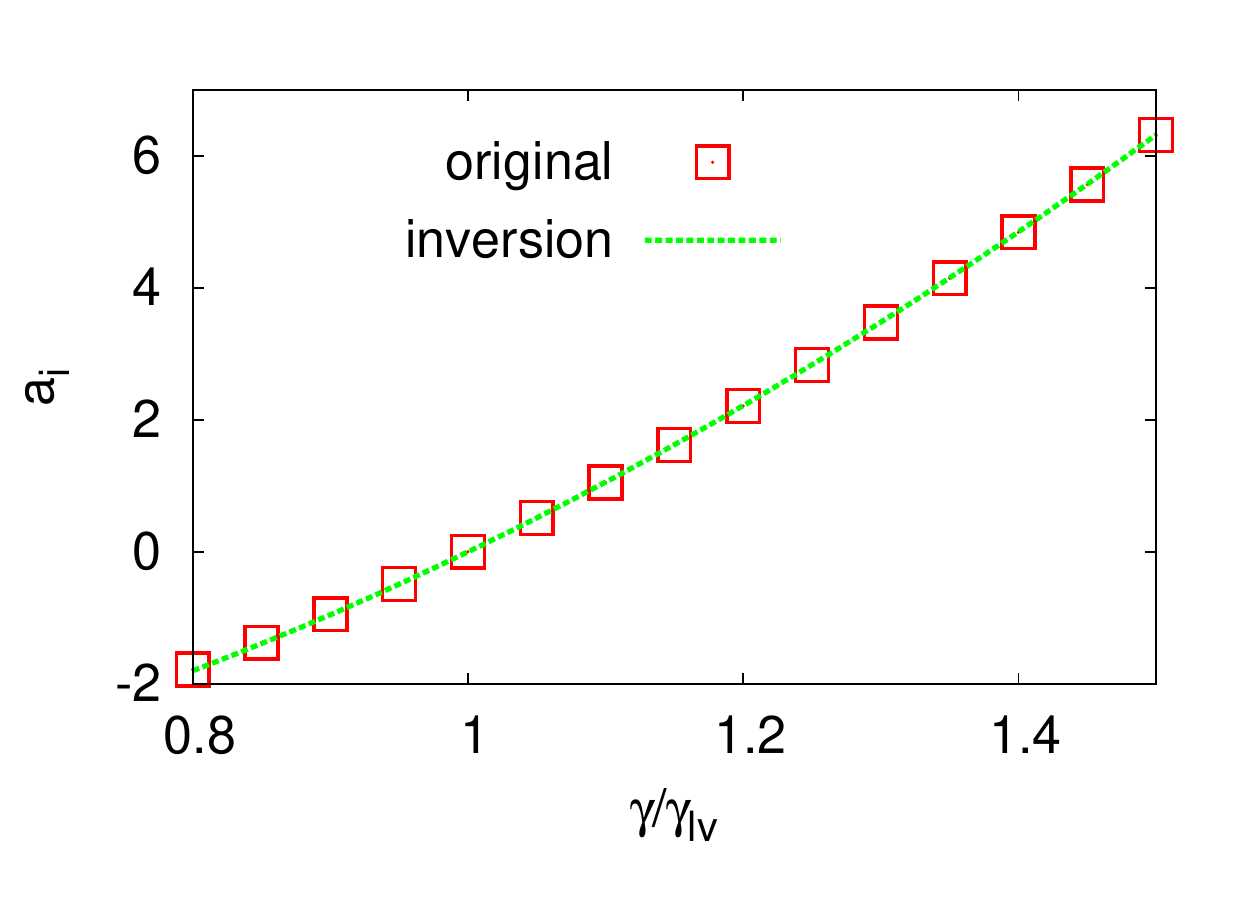}\\
  \caption{Comparison of free-energy barrier height parameter $a_i$ versus dimensionless interface energy
  $\gamma/\gamma_{lv}$ computed using the inversion formula Eq.~\ref{a_gamma_inversion} (green line),
  which predicts $a_i$ as a function of $\gamma/\gamma_{lv}$, and Eq. \ref{analytical_integral} (red square), which predicts $\gamma/\gamma_{lv}$ as a function of $a_i$. The inversion formula can be used to choose $a_i$ in the PF model to reproduce an arbitrary form of interface energy anisotropy.}
\label{a_gamma}
\end{center}
\end{figure}

In general, $a_i(\theta)$ needs to be computed to quantitatively reproduce a prescribed form of interface free-energy anisotropy.
For this, we start from the relation between
$a_i(\theta)$ and $\gamma_{jk}(\theta)$ given by
\begin{equation}
\frac{\gamma_{jk}(\theta)}{Wh}\equiv \tilde \gamma_{jk}
=2\sqrt{2}\int_0^1p(1-p)\sqrt{1+a_i(\theta)p(1-p)}dp,
\label{growth_app_facet_gamma}
\end{equation}
which can be reduced to
\begin{eqnarray}
& &\tilde \gamma_{jk} = 2\sqrt{2}\times\label{analytical_integral}\\
&
&\frac{2\sqrt{a_i}(4+3a_i)+(4+a_i)(3a_i-4)\cot^{-1}(2/\sqrt{a_i})}{64a_i^{3/2}}\nonumber,
\end{eqnarray}
by carrying out the integral. Eq.~\ref{analytical_integral} is a
transcendental equation and cannot be inverted analytically to find
$a_i$ as a function of $\tilde \gamma_{jk}$. However, a plot of
$a_i$ versus $\tilde \gamma_{jk}$ using
Eq.~\ref{analytical_integral} shows that the inverse function
$a_i(\tilde \gamma_{jk})$ is very accurately fitted over a wide
range of $a_i$ up to 100 by a simple quadratic polynomial.
\begin{equation}
a_i=B_0+B_1\tilde \gamma_{jk}+B_2\tilde
\gamma_{jk}^2,\label{a_gamma_inversion}
\end{equation}
with $B_0=-4.86349$, $B_1=-0.693313$ and $B_2=23.3564$. For the
$\gamma$ range we used in this work, accuracy of the quadratic
inversion formula is shown in Fig.~\ref{a_gamma}

\begin{figure}
  \begin{center}
  \includegraphics[width=80mm]{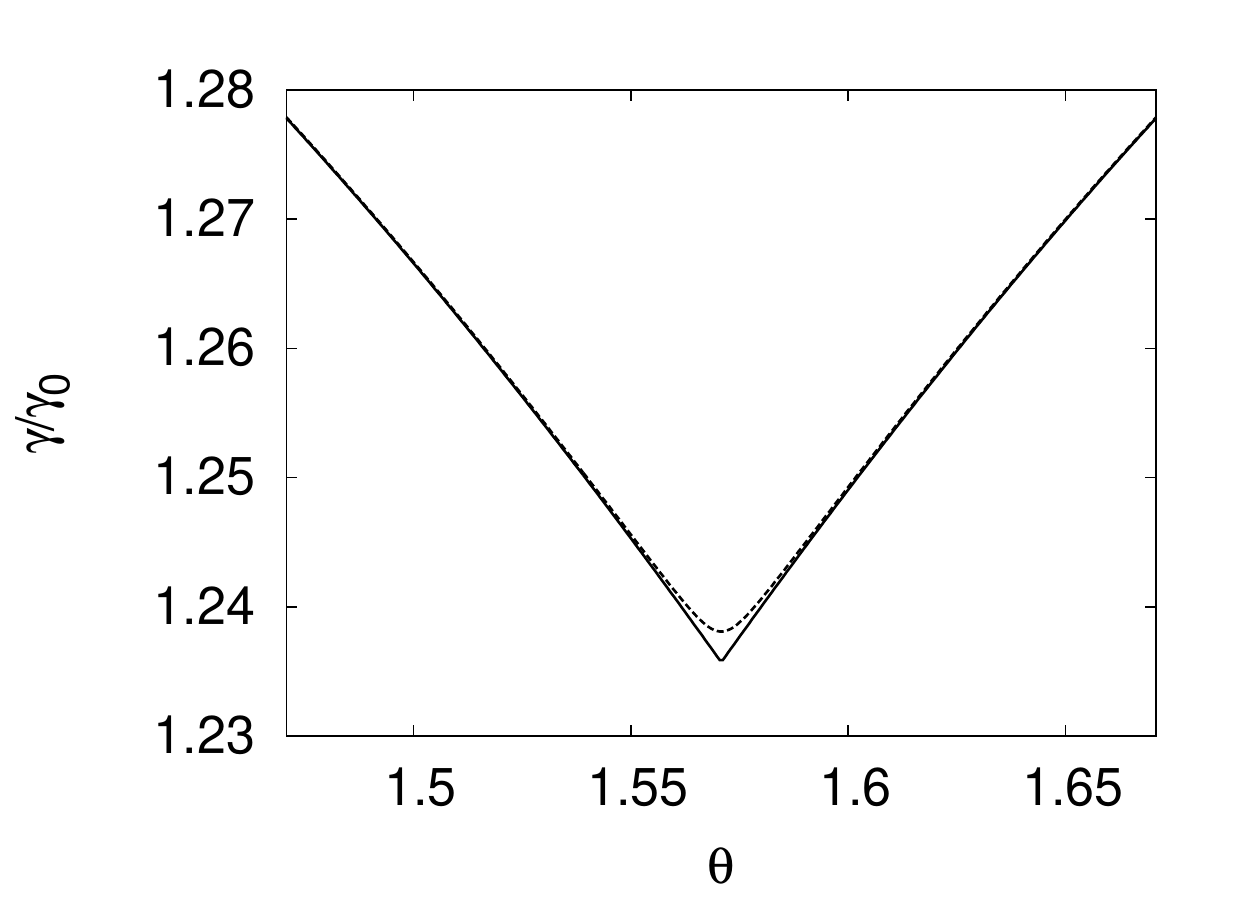}\\
  \caption{Regularization of the $\gamma$-plot near a cusp with $\delta_a=\delta_b=\sqrt{2}/6$ and $\epsilon=0.01$.
  The solid line is the original $\gamma$ plot given by Eq.~\ref{simple_gamma}.
  The dashed line is the regularized $\gamma$ plot given by Eq.~\ref{app_gamma_regular}.}
  \label{gamma_regular}
\end{center}
\end{figure}

The anisotropic surface energy in Eq.~\ref{simple_gamma} also needs
to be regularized since $d\gamma/d\theta$ becomes infinite at a sharp cusp. A simple regularized form is
\begin{equation}
\gamma(\theta)=\gamma^0[1+\delta_a\sqrt{\sin^2
2\theta+\epsilon^2}+\delta_b\sqrt{\cos^2
2\theta+\epsilon^2}],\label{app_gamma_regular}
\end{equation}
which is compared to the form of $\gamma$ with sharp cusps in Fig.~\ref{gamma_regular}.
The regularization parameter $\epsilon=0.01$ and interface cutoff
$W\epsilon_i=10^{-5}$ are used in all the numerical simulations with the
anisotropic model.

\end{document}